\documentclass[%
aip,
jcp,%
amsmath,amssymb,
preprint,%
]{revtex4-1}

\usepackage{graphicx}
\graphicspath{{./figures/}}
\usepackage{dcolumn}
\usepackage{bm}

\usepackage[version=3]{mhchem}
\usepackage{xspace}
\usepackage{siunitx}
\usepackage{hyperref}
\usepackage{braket}
\usepackage[nolist]{acronym}
\usepackage[inline]{enumitem}
\usepackage{algorithm}
\usepackage{algpseudocode}
\usepackage{multirow}

\usepackage[scale=0.85]{goldstone} 
\usepackage{simplewick}

\graphicspath{{./figures/}}

\newcommand{\psicode}{\href{http://www.psicode.org}{\textsc{Psi4}}\xspace}

\newcommand{\diff}{\mathop{}\!\mathrm{d}} 

\newcommand{\deriv}[3][]{
\frac{\diff^{#1}#2}{\diff #3^{#1}}}
\newcommand{\pderiv}[3][]{%
\frac{\partial^{#1}#2}
{\partial #3^{#1}}}
\newcommand{\vect}[1]{\bm{#1}} 
\newcommand{\mat}[1]{\bm{#1}}

\newcommand{\hairsp}{\hspace{1pt}}
\newcommand{\ie}{\textit{i.\hairsp{}e.}\xspace}

\newcommand{\eg}{\textit{e.\hairsp{}g.}\xspace}

\newcommand{\etal}{\textit{et al.}\xspace}

\newcommand{\Det}[1]{D_{\textbf{#1}}}
\newcommand{\refd}[0]{\ket{\Det{0}}}
\newcommand{\sdet}[1]{\ket{\Det{#1}}}
\newcommand{\ccwav}[0]{\ket{\mathrm{CC}}}

\newcommand{\commutator}[3][]{[ #2, #3 ]^{#1}}

\newcommand{\expT}[1]{\exp\left(#1\right)} 

\newcommand{\tamp}[2]{t_{#1}^{#2}}
\newcommand{\tampc}[1]{t_{\textbf{#1}}}

\newcommand{\excitor}[1]{\hat{\tau}_{\textbf{#1}}}

\newcommand{\BCHfirst}[2]{\commutator{#1}{#2}}
\newcommand{\BCHsecond}[2]{\frac{1}{2!}\commutator{\commutator{#1}{#2}}{#2}}
\newcommand{\BCHthird}[2]{\frac{1}{3!}\commutator{\commutator{\commutator{#1}{#2}}{#2}}{#2}}
\newcommand{\BCHfourth}[2]{\frac{1}{4!}\commutator{\commutator{\commutator{\commutator{#1}{#2}}{#2}}{#2}}{#2}}

\newcommand{\nestcommR}[3]{({#1}\underline{#2}^{#3})}

\newcommand{\oneBody}[2]{{h}_{#1}^{#2}}
\newcommand{\fock}[2]{{f}_{#1}^{#2}}
\newcommand{\twoBody}[4]{{g}_{#1 #2}^{#3 #4}}
\newcommand{\twoBodya}[4]{\bar{g}_{#1 #2}^{#3 #4}}

\newcommand{\iterate}[2]{{#1}^{[#2]}}
\DeclareMathOperator{\sgn}{sgn}
\newcommand{\order}[1]{\mathcal{O}\left(#1\right)}

\begin{document}

\title{Theory and Implementation of a Novel Stochastic Approach to Coupled Cluster}

\author{Charles J. C. Scott}
\affiliation{Department of Physics, King’s College London, Strand, London WC2R 2LS, United Kingdom}
\email{cjcargillscott@gmail.com}

\author{Roberto Di Remigio}
\affiliation{Hylleraas Centre for Quantum Molecular Sciences, Department of Chemistry, UiT The Arctic University of Norway, N-9037 Troms{\o}, Norway}
\affiliation{Department of Chemistry, Virginia Tech, Blacksburg, Virginia, United States}
\email{roberto.d.remigio@uit.no}

\author{T. Daniel Crawford}
\affiliation{Department of Chemistry, Virginia Tech, Blacksburg, Virginia, United States}
\affiliation{Molecular Sciences Software Institute, Blacksburg, Virginia, USA}

\author{Alex J. W. Thom}
\affiliation{Department of Chemistry, University of Cambridge, Cambridge, UK}

\begin{abstract}
  We present a detailed discussion of our novel \ac{diagCCMC} [Scott \emph{et al.} J. Phys. Chem. Lett. 2019, 10, 925].
  The \ac{diagCCMC} algorithm performs an imaginary-time propagation of the
  similarity-transformed coupled cluster Schr\"{o}dinger equation.
  Imaginary-time updates are computed by stochastic sampling of the coupled
  cluster vector function: each term is evaluated as a randomly realised diagram
  in the connected expansion of the similarity-transformed Hamiltonian.
  We highlight similarities and differences between deterministic and stochastic
  linked coupled cluster theory when the latter is re-expressed as a sampling of
  the diagrammatic expansion, and discuss details of our implementation
  that allow for a walker-less realisation of the stochastic sampling. Finally, we
  demonstrate that in the presence of locality, our algorithm can obtain a fixed errorbar per electron while only requiring an asymptotic computational
  effort that scales \emph{quartically} with system size, \emph{independently} of truncation level in \acl{CC} theory. 
  The algorithm only requires an asymptotic memory costs scaling linearly,
  as demonstrated previously. 
  These scaling reductions require no \emph{ad hoc} modifications to the approach.
\end{abstract}

\maketitle

\begin{acronym}
  \acro{LGPL}{GNU Lesser General Public Licence}
  \acro{WF}{wave function}
  \acro{WFT}{wave function theory}
  \acro{LHS}{left-hand side}
  \acro{CC}{coupled cluster}
  \acro{CCS}{coupled cluster with single substitutions}
  \acro{CCSD}{coupled cluster with single and double substitutions}
  \acro{CCSD(T)}{CCSD with perturbative triples correction}
  \acro{CCSDT}{coupled cluster with single, double and triple substitutions}
  \acro{CCSDTQ}{coupled cluster with up to quadruple  substitutions}
  \acro{CC2}{approximate coupled cluster singles and doubles}
  \acro{CC3}{approximate coupled cluster singles, doubles and triples}
  \acro{PT}{perturbation theory}
  \acro{MBPT}{many-body perturbation theory}
  \acro{MO}{molecular orbital}
  \acro{AO}{atomic orbital}
  \acro{MP}{M{\o}ller--Plesset}
  \acro{HF}{Hartree--Fock}
  \acro{QM}{quantum mechanics}
  \acro{QC}{quantum chemistry}
  \acro{RHS}{right-hand side}
  \acro{SCF}{self-consistent field}
  \acro{BCH}{Baker--Campbell--Hausdorff}
  \acro{MC}{Monte Carlo}
  \acro{FCI}{full configuration interaction}
  \acro{DMC}{diffusion Monte Carlo}
  \acro{QMC}{quantum Monte Carlo}
  \acro{VMC}{variational Monte Carlo}
  \acro{RI}{resolution-of-the-identity}
  \acro{FCIQMC}{full configuration interaction quantum Monte Carlo}
  \acro{SCCT}{stochastic coupled cluster theory}
  \acro{RDM}{reduced density matrix}
  \acrodefplural{RDM}[RDMs]{reduced density matrices}
  \acro{MSQMC}{Model Space Quantum Monte Carlo}
  \acro{CCMC}{coupled cluster Monte Carlo}
  \acro{diagCCMC}{diagrammatic coupled cluster Monte Carlo}
  \acro{RHF}{restricted Hartree--Fock}
  \acro{UHF}{unrestricted Hartree--Fock}
  \acro{ODE}{ordinary differential equation}
  \acro{EPV}{exclusion-principle violating}
\end{acronym}

\section{Introduction}

The goal of quantum chemistry is to provide accurate and cost-effective
methodologies for the solution of the molecular electronic Schr\"{o}dinger
equation.
One needs to be able not only to reproduce experimentally measurable
observables, but also to understand the microscopic origin of these measurements
and eventually predict and guide experiments.

Stochastic approaches to the solution of the Schr\"{o}dinger equation provide an
appealing alternative to deterministic strategies and a number of \ac{MC}
sampling methods have been continuously developed since the early days of
quantum chemistry.\cite{Foulkes2001-lh,Kolorenc2011}
At the cost of introducing statistical uncertainty in the results, \ac{QMC}
offers a low-scaling, parallelizable route to high-accuracy results. Despite the
favorable scaling and scalability, \ac{QMC} suffers from two well-known
problems. The statistical errorbar can be decreased, but at a very slow rate
with increasing length of the simulation, that is, a larger number of random
samples. Furthermore, for fermionic systems of interest in molecular electronic
structure theory, the nodal structure of the wavefunction needs to be fixed
\emph{a priori} to avoid collapse onto the bosonic ground state,\cite{Umrigar2015-fy,
Austin2012} introducing an uncontrolled approximation which thus far cannot be efficiently
relaxed to exactness. Despite this, their low polynomial scaling allows large-scale
applications, particularly within condensed matter systems where they can provide accurate
results while allowing extrapolation to remove finite-size errors.\citep{Foulkes2001-lh, Kolorenc2011, Hunt2018}

Within the deterministic realm, the two abovementioned problems do not appear.
No statistical uncertainity riddles the results and the methods are all
formulated in the appropriate Fock space, guaranteeing that the solution, while
approximate, is properly antisymmetrised. 
The \ac{CC} wavefunction Ansatz arguably provides the most effective framework
for accurate simulations of single-reference molecular systems. The \ac{CC}
model provides an exponential parametrization of the molecular electronic
wavefunction and enjoys a number of favorable properties.
It provides a systematic route towards the exact, \ac{FCI} solution, while
maintaining size-extensivity and -consistency of results at any truncation
level. Despite the exponential, nonlinear parametrization of the wavefunction,
the computational cost of \ac{CC} theory scales as a polynomial of system size,
albeit with potentially high values for the exponents.
Scaling and scalability are thus much less favorable than with stochastic
approaches: a number of approximations has to be introduced\cite{Pulay1983-ht,
Saebo1993-qx, Hampel1996-yy, Schutz2001-da, Neese2009-ch, Ziolkowski2010-oo,
Kristensen2011-ck, Hoyvik2012-td, Riplinger2013-mz, Riplinger2013-ue,
Eriksen2015-kw, Liakos2015-yw, Riplinger2016-zo, Pavosevic2016-kw,
Pavosevic2017-tl, Saitow2017-rr, Guo2018-bg, Yang2012-je, Schwilk2015-iv,
Ma2017-rl, Schwilk2017-zf, Ma2018-jb}
and many technical challenges need to be
surmounted.\cite{Matthews2018-yf,Solomonik2014-ss,Hartono2009-lv,Lewis2016-jw,Kats2013-zn,Lyakh2019-gm,Epifanovsky2013-te,Ibrahim2014-nx,Lyakh2018-wg}
In particular, while many high-performance implementations of \ac{CCSD} and
\ac{CCSD(T)} are nowadays available, the large gain in efficiency seen in local
theories has yet to be reproduced for higher truncation levels in the \ac{CC} hierarchy. 

With these considerations in mind, efforts in the past decade have been directed
at combining the best of both worlds into the formulation and implementation of
Fock-space \ac{QMC} methods.
The \ac{FCIQMC} was the first such method to be presented: the FCI secular problem
is solved as the dynamics of a population of signed particles.\cite{Booth2009,
Cleland2010, Petruzielo2012, Blunt2015-iq} This results in an exponentially
scaling algorithm, but with a dramatically reduced prefactor.
Building on this, some of us have further developed a similar projector \ac{MC}
algorithm to solve the \emph{unlinked} and \emph{linked} \ac{CC}
equations.\cite{Thom2010,Spencer2015a,Franklin2016,Scott2017,Neufeld2017}
These \ac{CC}\ac{MC} algorithms, implemented in the HANDE-QMC software
package,\cite{Spencer2019-tc} are fully general with respect to the excitation
level, allowing one to perform arbitrary order \ac{CC} simulations with a sparse
representation of the wavefunction.

We recently showed that neither \ac{FCIQMC} nor \ac{CC}\ac{MC} rigorously
fulfills size-extensivity for noninteracting systems.\cite{Scott2019-ge} Both
algorithms perform imaginary-time propagation of unlinked many-body
equations\cite{Harris2020-vu} which results in the unnecessary sampling of
zero-on-average terms. This unnecessary work negatively impacts the memory and
CPU costs of the simulation and is particularly severe for \ac{CC}\ac{MC}, as it
quickly precludes scaling to larger systems and/or higher orders of \ac{CC}
theory.
To remedy this situation, we put forth a \ac{MC} algorithm that performs the
imaginary-time propagation governed by the \emph{linked} \ac{CC} equations.
These are evaluated by random sampling of the connected terms in the
similarity-transformed Hamiltonian, conveniently represented as a diagrammatic
expansion. The diagCCMC algorithm restores size-extensivity and our preliminary
tests have shown how localization can be readily exploited without further
assumptions.
\footnote{ Let us note the existence of the diagrammatic \ac{MC} (DiagMC) method
in the quantum many-body literature.\cite{Van_Houcke2010-cz} Both DiagMC and
diagCCMC deal with integral equations by sampling in diagram space, but the
diagrams that are sampled are markedly different: quantum statistics models,
with denumerable, \emph{infinitely} many diagrams, and \ac{CC} wavefunctions,
with \emph{finitely} many diagrams. In addition, \emph{divergent} series might
arise in DiagMC requiring the stochastic realisation to handle the
resummation.\cite{Prokofev2007-pr} These differences lead to quite distinct
approaches to the sampling of terms. Despite the similar names, the two
techniques have fairly little in common.}

We should note that this is not the only avenue towards leveraging the benefits of
\ac{MC} sampling within the \ac{CC} approach. Deustua \etal have
shown how deterministic iterative \ac{CC} solvers can be seeded with amplitudes
from partially converged Fock-space \ac{QMC} results. Combined with moment
expansion corrections,\cite{Deustua2019-rw, Deustua2018-bu, Deustua2017-ks} this
approach is a powerful technique, enabling access to higher levels of \ac{CC}
theory at reduced cost.
However, the high computational scaling of the \ac{QMC} methods used to determine important higher-level amplitudes will eventually dominate the overall computational cost of this approach, and thus our work provides a complementary solution.

In this work, we will first describe in detail the theoretical framework on
which our diagCCMC algorithm rests.
Section~\ref{sec:background} summarises \ac{CC} theory, with particular emphasis
on its diagrammatic formulation.
In Section~\ref{sec:stochastic-cc} we present a derivation of the imaginary-time
update step and its usage in the \ac{CC}\ac{MC} and diagCCMC algorithms.
We will then discuss the structure of the implemented algorithm and 
highlight differences and similarities to a deterministic implementation of \ac{CC} theory.

\section{Background Theory}\label{sec:background}
\subsection{Notation}\label{sec:notation}

We will use the tensor notation for second quantization.\cite{Harris1981-pb,Kutzelnigg1982-oq,Kutzelnigg1997-vt}
We denote elementary, anticommuting fermion creation and annihilation operators as:
\begin{alignat}{2}
  \hat{a}_{p}^\dagger = \hat{a}^p \equiv \hat{p}^+, &\quad 
  \hat{a}_{p} \equiv \hat{p}^-.
\end{alignat}
A $k$-electron excitation operator with respect to the physical vacuum
($\ket{\text{vac}}$) is the product of $k$ creation and $k$ annihilation
operators. In tensor notation:
\begin{equation}
  \hat{a}_{r_1r_2\ldots r_k}^{s_1s_2\ldots s_k} = \hat{s}_1^+\hat{s}_2^+\ldots \hat{s}_k^+\hat{r}_k^-\ldots \hat{r}_2^- \hat{r}_1^-,
\end{equation}
such excitation operators are particle number-conserving.
Explicitly, the one- and two-electron substitutions are:
\begin{alignat}{2}
  \hat{a}_{q}^{p} = \hat{p}^+\hat{q}^-, &\quad 
  \hat{a}_{rs}^{pq} &= \hat{p}^+\hat{q}^+\hat{s}^- \hat{r}^-
\end{alignat}

The Born--Oppenheimer, molecular electronic Hamiltonian is then expressed as:
\begin{equation}\label{eq:molecular-hamiltonian}
  \hat{H} = \sum_{pq} \oneBody{p}{q}\hat{a}_{q}^{p} + \frac{1}{2}\sum_{pqrs}\twoBody{p}{q}{r}{s}\hat{a}_{rs}^{pq} 
          = \sum_{pq} \oneBody{p}{q}\hat{a}_{q}^{p} + \frac{1}{4}\sum_{pqrs}\twoBodya{p}{q}{r}{s}\hat{a}_{rs}^{pq}
\end{equation}
where the integrals are given in an orthonormal basis of one-electron spin-orbitals:
\begin{subequations}
  \begin{align}
    \oneBody{p}{q}   &= \int\diff\mathbf{x} \phi^{*}_p(\mathbf{x}) \left( -\frac{1}{2}\nabla^2 + V_{\mathrm{eN}} \right) \phi_q(\mathbf{x}) \\
    \twoBody{p}{q}{r}{s} &=
                           \int\diff\mathbf{x}\int\diff\mathbf{x}^{\prime}\frac{\phi^{*}_p(\mathbf{x})\phi^{*}_q(\mathbf{x}^{\prime})\phi_r(\mathbf{x})\phi_s(\mathbf{x}^{\prime})}{|\mathbf{r}_1 - \mathbf{r}_2|} = \braket{pq | rs} \\
   \twoBodya{p}{q}{r}{s} &= g_{pq}^{rs} -  g_{pq}^{sr}= \braket{pq || rs}.
  \end{align}
\end{subequations}

For single-reference theories, it is more convenient to work in terms of the
Fermi, rather than the physical, vacuum state.
Our Fermi vacuum will be a single-determinant reference function $\refd$ for an
$N$-electron system with $2M$ spin-orbitals.

Occupied one-particle states in $\refd$, $i_1,i_2,\ldots,i_N$, will be referred
to as \emph{hole} states, whereas virtual one-particle states, $a_{N+1},
a_{N+2},\ldots$ will be referred to as \emph{particle} states.
A normal-ordered, $k$-electron substitution operator will be denoted as:
\begin{equation}
  \hat{e}_{r_1r_2\ldots r_k}^{s_1s_2\ldots s_k}
\end{equation}
Using Wick's theorem,\cite{Wick1950-iy, Kutzelnigg1997-vt} these operators can
be rewritten as a finite sum of subsets of permutations of elementary operators
times contractions. The latter are elements of $k$-electron \aclp{RDM}\acused{RDM}
($k$-\acp{RDM}):
\begin{equation}
  \label{eq:k-rdm}
  \gamma_{r_1r_2\ldots r_k}^{s_1s_2\ldots s_k} \equiv \Braket{\Det{0} | \hat{a}_{r_1r_2\ldots r_k}^{s_1s_2\ldots s_k} | \Det{0}},
\end{equation}
Thus, the normal-ordered one- and two-electron substitutions are:\footnote{For a
  single-determinant reference function one has: $\gamma_{r_1}^{s_1} = \delta_{r_1}^{s_1}$ and
  $\gamma_{r_1r_2}^{s_1s_2} = \delta_{r_1}^{s_1}\delta_{r_2}^{s_2} - \delta_{r_1}^{s_2}\delta_{r_2}^{s_1}$.}
\begin{alignat}{2}
  \hat{e}_{r_1}^{s_1} = \hat{a}_{r_1}^{s_1} - \gamma_{r_1}^{s_1},\quad
  \hat{e}_{r_1r_2}^{s_1s_2} = \hat{a}_{r_1r_2}^{s_1s_2}
  - (\gamma_{r_1}^{s_1} \hat{a}_{r_2}^{s_2} + \gamma_{r_2}^{s_2} \hat{a}_{r_1}^{s_1} 
  - \gamma_{r_2}^{s_1} \hat{a}_{r_1}^{s_2}  - \gamma_{r_1}^{s_2} \hat{a}_{r_2}^{s_1}  - \gamma_{r_1r_2}^{s_1s_2}).
\end{alignat}

Imposing normal ordering on the molecular Hamiltonian we obtain:
\begin{equation}\label{eq:molecular-hamiltonian-N}
  \hat{H}_{\mathrm{N}} = \hat{F} + \hat{\Phi}
  = \sum_{pq} \fock{p}{q}\hat{e}_{q}^{p} + \frac{1}{4}\sum_{pqrs}\twoBodya{p}{q}{r}{s}\hat{e}_{rs}^{pq}
  = \hat{H} - E_{\mathrm{ref}}, 
\end{equation}
where the energy of the reference determinant, the one-body Fock operator, and
the two-body fluctuation potential appear:
\begin{equation}
  \begin{aligned}
    E_{\mathrm{ref}}
    &= \braket{\Det{0} | \hat{H} | \Det{0}}
    = \sum_{i} \oneBody{i}{i} + \frac{1}{2} \sum_{ij}\twoBodya{i}{j}{i}{j} \\
    \fock{p}{q} &= \oneBody{p}{q} + \sum_{i}\twoBodya{p}{i}{q}{i}
  \end{aligned}
\end{equation}
We will use the symbol $\excitor{k}$ for pure excitation operators, or
\emph{excitors}. These are $k$-electron substitutions between hole
and particle states in the reference determinant and thus particle number- and charge-conserving.

Since the $k$-\acp{RDM} in \eqref{eq:k-rdm} are zero whenever any of the indices, upper or lower, refers
to a particle state, the excitors are automatically normal-ordered:
\begin{equation}
  \begin{aligned}
  \excitor{k} &= \hat{a}_{i_1i_2\ldots i_k}^{a_1a_2\ldots a_k}
  = \hat{e}_{i_1i_2\ldots i_k}^{a_1a_2\ldots a_k} \\
  &= \hat{a}_1^+\hat{a}_2^+\ldots \hat{a}_k^+\hat{\imath}_k^-\ldots \hat{\imath}_2^- \hat{\imath}_1^-
  = \hat{a}_1^+\hat{\imath}_1^-\hat{a}_2^+\ldots \hat{a}_k^+\hat{\imath}_k^-.
  \end{aligned}
\end{equation}
Here we have introduced the multi-index $\textbf{k} = \left[^{a_1 a_2\ldots a_k}_{i_1 i_2\ldots i_k}\right]$
to compactly represent the $k$-electron
substitution effected by the excitor, which is, up to a phase, a $k$-excited determinant:
\begin{equation}
  \sdet{k} = \excitor{k}\refd \propto \left|^{a_1 a_2\ldots a_k}_{i_1 i_2\ldots i_k}\right\rangle
\end{equation}
A summary of our notation can be found in Table~\ref{tab:nomenclature}.

\begin{table}
  \caption{Overview of notation and nomenclature.}
  \label{tab:nomenclature}
 \centering
 \begin{ruledtabular}
  \begin{tabular}{r l}
    Symbol & Short description \\
    $\refd$ & The reference Slater determinant \\
    $\hat{a}^p$ & fermion creation operator \\
    $\hat{a}_{p}$ & fermion annihilation operator \\
    $\hat{a}_{r_1r_2\ldots r_k}^{s_1s_2\ldots s_k}$ & $k$-electron excitation operator with respect to the physical vacuum \\
    $\hat{e}_{r_1r_2\ldots r_k}^{s_1s_2\ldots s_k}$ & $k$-electron excitation operator, normal-ordered with respect to the reference \\
    $\textbf{j}, \textbf{k}, \ldots$ & Replacement multi-indices, \eg $\textbf{k} = \left[^{a_1 a_2\ldots a_k}_{i_1 i_2\ldots i_k}\right]$ \\
    $\sdet{k}$ & The $\textbf{k}$-th replacement excited determinant \\ 
    $p, q, r, s, \ldots$ & General spin-orbital indices \\
    $i_1, i_2, \ldots, i_k$ & Hole spin-orbitals in $\refd$ \\
    $a_1, a_2, \ldots, a_k$ & Particle spin-orbitals in $\refd$ \\
    $\excitor{k}$ & \textbf{Excitor} for the $\textbf{k}$-th replacement \\
    $\tampc{k}$ & \textbf{Cluster amplitude} for the $\textbf{k}$-th replacement \\
    $\tampc{k}\excitor{k}$ & \textbf{Connected} (\textbf{Non-composite}) cluster \\ 
    $\frac{1}{2!}\tampc{k}\tampc{l}\excitor{k}\excitor{l}$ & \textbf{Disconnected} (\textbf{Composite}) cluster \\
  \end{tabular}
  \end{ruledtabular}
\end{table}

\subsection{The coupled cluster Ansatz}\label{sec:general-cc-info}

The coupled-cluster wavefunction is parametrized as an exponential
transformation of a reference single-determinant wavefunction $\refd$:
\begin{equation}\label{eq:standard-cc-wav}
  \ccwav = \expT{\hat{T}} \refd,
\end{equation}
where the cluster operator $\hat{T}$ is given as a sum of second-quantised excitation operators:
\begin{equation}\label{eq:cluster-op}
  \hat{T} = \sum_{m} \hat{T}_{m},
\end{equation}
with the $m$-th order cluster operators expressed as sums of excitors weighted
by the corresponding \emph{cluster amplitudes}:
\begin{equation}\label{eq:cluster-k}
  \hat{T}_{m} = \sum_{\textbf{k} \in m^{\mathrm{th}} \text{replacements}} \tampc{k}\excitor{k}
  = \frac{1}{(k!)^2} \sum_{\substack{a_1, a_2, \ldots, a_k \\ i_1, i_2, \ldots, i_k}}
  \tamp{a_1a_2\ldots a_k}{i_1i_2\ldots i_k}
  \hat{e}_{i_1i_2\ldots i_k}^{a_1a_2\ldots a_k}.
\end{equation}
Note that in the tensor notation adopted, upper and lower indices
of the cluster amplitudes appear reversed with respect to other conventions.

The \ac{CC} correlation energy is the right eigenvalue of the Schr\"{o}dinger
equation for the normal-ordered Hamiltonian in~(\ref{eq:molecular-hamiltonian-N}): 
\begin{equation}\label{eq:unlinked-schrodinger}
  \hat{H}_{\mathrm{N}}\ket{\mathrm{CC}} = \Delta E_{\mathrm{CC}}\ket{\mathrm{CC}}.
\end{equation}
This equation is solved by performing a similarity transformation of the
Hamiltonian:
\begin{equation}\label{eq:linked-schrodinger}
  \expT{-\hat{T}} \hat{H} \expT{\hat{T}} \refd = \bar{H}\refd = \Delta E_{\mathrm{CC}} \refd
\end{equation}
and then projecting onto the excitation manifold $\lbrace\sdet{j}\rbrace$:
\begin{subequations}\label{eq:linked-cc}
  \begin{align}
    \braket{\Det{0} | \bar{H} | \Det{0}} &= \Delta E_\mathrm{CC} \label{eq:linked-cc-energy} \\
    \braket{\Det{k} | \bar{H} | \Det{0}} &= \omega_{\textbf{k}}(\vect{t}). \label{eq:linked-cc-amplitudes}
  \end{align}
\end{subequations}
The second equation defines the \ac{CC} residual
$\omega_{\textbf{k}}(\vect{t})$, which is zero at a solution of the nonlinear linked equations.\cite{Piecuch2000-nd}
Whereas~\eqref{eq:unlinked-schrodinger} and~\eqref{eq:linked-cc-energy} can be proved
to be identical,\cite{Helgaker2000-yb} the \emph{linked} formulation in the
latter is size-extensive order-by-order and term-by-term.
For notational convenience, we have dropped the subscript $\mathrm{N}$ for the Hamiltonian.
The similarity-transformed Hamiltonian $\bar{H}$ can be expanded into a \ac{BCH}
commutator series:
\begin{alignat}{2}\label{eq:bch}
  \expT{-\hat{T}} \hat{H} \expT{\hat{T}} = \bar{H} = \sum_{n \geq 0} \frac{1}{n!} \nestcommR{\hat{H}}{\hat{T}}{n},\quad&
    \nestcommR{\hat{H}}{\hat{T}}{} \stackrel{\textrm{def}}{=} \BCHfirst{\hat{H}}{\hat{T}}.
\end{alignat}
For the molecular Hamiltonian in equation \eqref{eq:molecular-hamiltonian}, at
most two-body operators are involved. Hence, regardless of the truncation level
in the cluster operator $\hat{T}$, the expansion truncates at the four-fold nested
commutator:
\begin{equation}\label{eq:H-commutator-expansion}
\begin{aligned}
  \bar{H} &= \sum^4_{n = 0} \frac{1}{n!} \nestcommR{\hat{H}}{\hat{T}}{n} \\
  &= \hat{H} + \BCHfirst{\hat{H}}{\hat{T}} + \BCHsecond{\hat{H}}{\hat{T}} \\
&+ \BCHthird{\hat{H}}{\hat{T}} + \BCHfourth{\hat{H}}{\hat{T}},
\end{aligned}
\end{equation}
showing that only finitely many terms are included in equation~\eqref{eq:linked-cc-amplitudes}.
Despite the fact that $\bar{H}$ is no longer Hermitian, the linked formulation
is still more advantageous than the unlinked formulation.\cite{Helgaker2000-yb,Shavitt2009-mr,Crawford2000}

Since all excitors are normal-ordered and commuting, Wick's
theorem\cite{Wick1950-iy, Crawford2000, Shavitt2009-mr} lets us
reduce the Hamiltonian-excitor products to only those terms which are
\emph{connected} (in the diagrammatic sense). Excitors will only appear to the \emph{right} of the
Hamiltonian and only terms where each excitor shares at least one index with
the Hamiltonian will be nonzero:
\begin{equation}\label{eq:H-connected-expansion}
  \bar{H} =
  (\hat{H}\expT{\hat{T}})_{\mathrm{c}} =
  \hat{H}
  + (\hat{H}\hat{T})_{\mathrm{c}}
  + \frac{1}{2!}(\hat{H}\hat{T}\hat{T})_{\mathrm{c}}
  + \frac{1}{3!}(\hat{H}\hat{T}\hat{T}\hat{T})_{\mathrm{c}}
  + \frac{1}{4!}(\hat{H}\hat{T}\hat{T}\hat{T}\hat{T})_{\mathrm{c}}.
\end{equation}
The requirement of shared indices between the Hamiltonian and cluster
coefficients enables the resulting equations to be solved via a series of tensor
contractions: a process highly amenable to rapid evaluation on conventional
computing architectures,\cite{Crawford2019-yz} but non-trivial to parallelise.\cite{Matthews2018-yf, Solomonik2014-ss, Ibrahim2014-nx}

\subsection{Diagrammatic representation}\label{sec:diagrammatic-representation}

The algebraic derivation of the linked \ac{CC} equations to a general
truncation level from equation~\eqref{eq:linked-cc-amplitudes} is lengthy and error prone.
A diagrammatic representation can be effectively used to generate all unique
terms in the equations.\cite{Kucharski1986-sz, Bartlett2007-mz, Shavitt2009-mr}
Normal-ordering and application of Wick's theorem are key to these developments.
The normal-ordered Hamiltonian features 13 \emph{interaction vertices}, 4 coming
from the Fock operator:
\begin{equation}
  F = \sum_{a_1a_2}\fock{a_1}{a_2}\hat{e}_{a_2}^{a_1}
    + \sum_{i_1i_2}\fock{i_1}{i_2}\hat{e}_{i_2}^{i_1} 
    + \sum_{i_1a_1}\fock{i_1}{a_1}\hat{e}_{a_1}^{i_1}
    + \sum_{a_1i_1}\fock{a_1}{i_1}\hat{e}_{i_1}^{a_1},
\end{equation}
and 9 from the fluctuation potential:
\begin{equation}
  \begin{aligned}
  \Phi &=
    \frac{1}{4}\sum_{\substack{a_1a_2 \\ a_3a_4}}\twoBodya{a_1}{a_2}{a_3}{a_4}\hat{e}_{a_3a_4}^{a_1a_2}
  + \frac{1}{4}\sum_{\substack{i_1i_2 \\ i_3i_4}}\twoBodya{i_1}{i_2}{i_3}{i_4}\hat{e}_{i_3i_4}^{i_1i_2}
  +            \sum_{\substack{a_1i_1 \\ a_2i_2}}\twoBodya{a_1}{i_1}{a_2}{i_2}\hat{e}_{a_2i_2}^{a_1i_1} \\
  &+ \frac{1}{2}\sum_{\substack{a_1a_2 \\ a_3i_1}}\twoBodya{a_1}{a_2}{a_3}{i_1}\hat{e}_{a_3i_1}^{a_1a_2}
   + \frac{1}{2}\sum_{\substack{i_1a_1 \\ i_2i_3}}\twoBodya{i_1}{a_1}{i_2}{i_3}\hat{e}_{i_2i_3}^{i_1a_1}
   + \frac{1}{2}\sum_{\substack{a_1i_1 \\ a_2a_3}}\twoBodya{a_1}{i_1}{a_2}{a_3}\hat{e}_{a_2a_3}^{a_1i_1} \\
  &+ \frac{1}{2}\sum_{\substack{i_1i_2 \\ i_3a_1}}\twoBodya{i_1}{i_2}{i_3}{a_1}\hat{e}_{i_3a_1}^{i_1i_2}
   + \frac{1}{4}\sum_{\substack{i_1i_2 \\ a_1a_2}}\twoBodya{i_1}{i_2}{a_1}{a_2}\hat{e}_{a_1a_2}^{i_1i_2}
   + \frac{1}{4}\sum_{\substack{a_1a_2 \\ i_1i_2}}\twoBodya{a_1}{a_2}{i_1}{i_2}\hat{e}_{i_1i_2}^{a_1a_2}
  \end{aligned}
\end{equation}
Each of these vertices can be characterized by an integer representing their
excitation level (0, $\pm 1$, $\pm 2$) and by a sign sequence encoding the
pattern of open particle ($+$) and hole ($-$) lines \emph{below} the interaction vertex, see Table~\ref{tab:H-vertices}.
Cluster operators can be classified similarly in terms of their excitation level
(any integer $\geq 1$) and their sign sequence.

\begin{table}
 \caption{The thirteen interaction vertices of the normal-ordered Hamiltonian
   with corresponding matrix elements, excitation levels, and Kucharski--Bartlett
   sign sequences.\cite{Kucharski1986-sz}}\label{tab:H-vertices}
 \centering
 \begin{ruledtabular}
   \begin{tabular}{l c c c c}
      & Vertex & Matrix element & Excitation level & Sign sequence \\
     1 &
$\diagram{
  \draw (-0.5,0) node[circlex] (f) {} -- (0,0) node[dot=white] (f1) {};
  \draw[->-] (f1) to node[right] {\tiny $a_{1}$} ++(0,+0.5);
  \draw[-<-] (f1) to node[right] {\tiny $a_{2}$} ++(0,-0.5);
}$
     & $\fock{a_1}{a_2}$ & 0 & $+$ \\[.5cm]
     2 &
         $\diagram{
         \draw (-0.5,0) node[circlex] (f) {} -- (0,0) node[dot=white] (f1) {};
         \draw[-<-] (f1) to node[right] {\tiny $i_{2}$} ++(0,+0.5);
         \draw[->-] (f1) to node[right] {\tiny $i_{1}$} ++(0,-0.5);
         }$
     & $\fock{i_1}{i_2}$ & 0 & $-$ \\[.5cm]
     3 &
$\diagram{
  \draw (-0.5,0) node[circlex] (f) {} -- (0,0) node[dot=white] (f1) {};
  \draw[->-] (f1) to node[left] {\tiny $i_{1}$} ++(-0.25,-0.5);
  \draw[-<-] (f1) to node[right] {\tiny $a_{1}$} ++(0.25,-0.5);
}$
               & $\fock{i_1}{a_1}$ & -1 & $+-$ \\[.5cm] 
     4 &
         $\diagram{
         \interaction{2}{g}{(0,0)}{dot=white}{flexdotted};
         \draw[->-] (g1) to node[left] {\tiny $a_{1}$} ++(0,+0.5);
         \draw[-<-] (g1) to node[left] {\tiny $a_{3}$}++(0,-0.5);
         \draw[->-] (g2) to node[left] {\tiny $a_{2}$}++(0,+0.5);
         \draw[-<-] (g2) to node[left] {\tiny $a_{4}$}++(0,-0.5);
         }$
               & $\twoBodya{a_1}{a_2}{a_3}{a_4}$ & 0 & $++$ \\[.5cm]
     5 &
$\diagram{
  \interaction{2}{g}{(0,0)}{dot=white}{flexdotted};
  \draw[-<-] (g1) to node[left] {\tiny $i_{3}$} ++(0,+0.5);
  \draw[->-] (g1) to node[left] {\tiny $i_{1}$}++(0,-0.5);
  \draw[-<-] (g2) to node[left] {\tiny $i_{4}$}++(0,+0.5);
  \draw[->-] (g2) to node[left] {\tiny $i_{2}$}++(0,-0.5);
}$
               & $\twoBodya{i_1}{i_2}{i_3}{i_4}$ & 0 & $--$ \\[.5cm] 
     6 &
         $\diagram{
         \interaction{2}{g}{(0,0)}{dot=white}{flexdotted};
         \draw[->-] (g1) to node[left] {\tiny $a_{1}$} ++(0,+0.5);
         \draw[-<-] (g1) to node[left] {\tiny $a_{2}$}++(0,-0.5);
         \draw[-<-] (g2) to node[left] {\tiny $i_{2}$}++(0,+0.5);
         \draw[->-] (g2) to node[left] {\tiny $i_{1}$}++(0,-0.5);
         }$
               & $\twoBodya{a_1}{i_1}{a_2}{i_2}$ & 0 & $+-$ \\[.5cm]  
     7 &
         $\diagram{
         \interaction{2}{g}{(0,0)}{dot=white}{flexdotted};
         \draw[->-] (g1) to node[left] {\tiny $a_{1}$} ++(0,+0.5);
         \draw[-<-] (g1) to node[left] {\tiny $a_{3}$}++(0,-0.5);
         \draw[-<-] (g2) to node[left] {\tiny $i_{1}$}++(-0.25,+0.5);
         \draw[->-] (g2) to node[right] {\tiny $a_{2}$}++(0.25,+0.5);
         }$
               & $\twoBodya{a_1}{a_2}{a_3}{i_1}$ & +1 & $+$ \\[.5cm]  
     8 &
$\diagram{
  \interaction{2}{g}{(0,0)}{dot=white}{flexdotted};
  \draw[-<-] (g1) to node[left] {\tiny $i_{2}$} ++(0,+0.5);
  \draw[->-] (g1) to node[left] {\tiny $i_{1}$}++(0,-0.5);
  \draw[-<-] (g2) to node[left] {\tiny $i_{3}$}++(-0.25,+0.5);
  \draw[->-] (g2) to node[right] {\tiny $a_{1}$}++(0.25,+0.5);
}$
     & $\twoBodya{i_1}{a_1}{i_2}{i_3}$ & +1 & $++-$ \\[.5cm]  
     9 &
$\diagram{
  \interaction{2}{g}{(0,0)}{dot=white}{flexdotted};
  \draw[->-] (g1) to node[left] {\tiny $a_{1}$} ++(0,+0.5);
  \draw[-<-] (g1) to node[left] {\tiny $a_{2}$}++(0,-0.5);
  \draw[->-] (g2) to node[left] {\tiny $i_{1}$}++(-0.25,-0.5);
  \draw[-<-] (g2) to node[right] {\tiny $a_{3}$}++(0.25,-0.5);
}$
     & $\twoBodya{a_1}{i_1}{a_2}{a_3}$ & -1 & $-$ \\[.5cm]  
     10 &
$\diagram{
  \interaction{2}{g}{(0,0)}{dot=white}{flexdotted};
  \draw[-<-] (g1) to node[left] {\tiny $i_{3}$} ++(0,+0.5);
  \draw[->-] (g1) to node[left] {\tiny $i_{1}$}++(0,-0.5);
  \draw[->-] (g2) to node[left] {\tiny $i_{2}$}++(-0.25,-0.5);
  \draw[-<-] (g2) to node[right] {\tiny $a_{1}$}++(0.25,-0.5);
}$
     & $\twoBodya{i_1}{i_2}{i_3}{a_1}$ & -1 & $+--$ \\[.5cm]  
     11 &
$\diagram{
  \interaction{2}{g}{(0,0)}{dot=white}{flexdotted};
  \draw[->-] (g1) to node[left] {\tiny $i_{1}$} ++(-0.25,-0.5);
  \draw[-<-] (g1) to node[right] {\tiny $a_{1}$}++(0.25,-0.5);
          \draw[->-] (g2) to node[left, label={[xshift=-0.1cm, yshift=-0.55cm] {\tiny $i_{2}$}}] {}++(-0.25,-0.5);
          \draw[-<-] (g2) to node[right] {\tiny $a_{2}$}++(0.25,-0.5);
          }$
               & $\twoBodya{i_1}{i_2}{a_1}{a_2}$ & -2 & $++--$ \\[.5cm]  
     12 &
          $\diagram{
          \draw (-0.5,0) node[circlex] (f) {} -- (0,0) node[dot=white] (f1) {};
          \draw[-<-] (f1) to node[left] {\tiny $i_{1}$} ++(-0.25,0.5);
          \draw[->-] (f1) to node[right] {\tiny $a_{1}$} ++(0.25,0.5);
          }$
               & $\fock{a_1}{i_1}$ & +1 & $0$ \\[.5cm]  
     13 &
          $\diagram{
          \interaction{2}{g}{(0,0)}{dot=white}{flexdotted};
          \draw[-<-] (g1) to node[left] {\tiny $i_{1}$} ++(-0.25,0.5);
          \draw[->-] (g1) to node[right, label={[xshift=0.2cm, yshift=-0.15cm] {\tiny $a_{1}$}}] {}++(0.25,0.5);
  \draw[-<-] (g2) to node[left] {\tiny $i_{2}$}++(-0.25,0.5);
  \draw[->-] (g2) to node[right] {\tiny $a_{2}$}++(0.25,0.5);
}$
     & $\twoBodya{a_1}{a_2}{i_1}{i_2}$ & +2 & $0$ \\
  \end{tabular}
 \end{ruledtabular}
\end{table}

For any given excitation level in the allowed manifold (up to double excitations
for CCSD, triple excitations for CCSDT, and so forth), the diagrammatic
generation of the corresponding \ac{CC} equations proceeds \emph{via} these steps:
\begin{enumerate}
  \item At the bottom, we draw a combination of at most four excitors.
  \item At the top, we draw a Hamiltonian vertex. The valid vertices are limited
    by two requirements:
    \begin{enumerate*}[label=\emph{\alph*})]
    \item the final diagram be connected and
    \item the overall excitation level of the projection manifold.
    \end{enumerate*}
  \item We pair the Hamiltonian vertex and excitor(s) sign sequences in all
    distinct ways to generate the sign sequences for all unique diagrams. The sign
    sequence encodes the diagram topology and ensuing contraction pattern.
  \item We read the algebraic expression for the
    corresponding term in the \ac{CC} equations off from the generated diagrams.
    The rules of interpretation
    associate target indices to the external (open) lines and dummy summation
    indices to the internal lines, Hamiltonian matrix elements to the
    interaction vertices and products of amplitudes to the excitor vertices.
    Topological and permutational symmetries are taken into account by similar
    simple rules.\cite{Kucharski1986-sz, Crawford2000, Shavitt2009-mr} 
\end{enumerate}

The rules for generating and interpreting diagrams as algebraic expressions are
independent of the \ac{CC} truncation order and can be encoded into a computer
program.\cite{Harris1999-gf, Crawford2000, Kallay2001-yv, Kallay2004-ug,
Kallay2003-qi, Kallay2004-fv, Lyakh2005-do}
However, a proper factorization of intermediates is essential to achieve
acceptable time to solution and memory requirements.\cite{Kallay2001-yv}

\section{Stochastic realizations of coupled cluster theory}\label{sec:stochastic-cc}

The solution of the \ac{CC} equations can be achieved by means of stochastic
algorithms. This stochastic realization is, however, not unique, and multiple
algorithms have been put forward in the literature.\cite{Franklin2016,Spencer2015a,Thom2010} 
All these different realizations are based on reformulating the time-dependent
Schr\"{o}dinger equation in imaginary-time. The corresponding diffusion-like
equation can be solved by repeated application of an approximate propagator on a
trial state. Employing a Fock space representation circumvents the fermion sign
problem, without the need for fixing the nodes \emph{a priori}.\cite{Spencer2012-pw}

\subsection{The imaginary-time propagation}

After performing a Wick rotation $\tau\leftarrow \mathrm{i}t$ to imaginary time,
the time-dependent \ac{CC} Schr\"{o}dinger equation reads as:\cite{Pigg2012-gd,Ten-No2017-zx}
\begin{equation}\label{eq:cc-imaginary-time-schro}
  \deriv{}{\tau}\left[\expT{\hat{T}(\tau)}\refd\right] = - \hat{H}\expT{\hat{T}(\tau)}\refd.
\end{equation}
The $\tau$-derivative on the left-hand side is (see Appendix \ref{app:expo-derivative}):
\begin{equation}
  \deriv{}{\tau}\expT{\hat{T}}
  =
  \expT{\hat{T}}
  \left\lbrace \sum_{l \geq 0} \frac{1}{(l+1)!} \nestcommR{\dot{\hat{T}}}{\hat{T}}{l} \right\rbrace.
\end{equation}
Excitation operators are assumed time-independent:
\begin{equation}
  \dot{\hat{T}}(\tau) = \sum_{\textbf{k}}\dot{t}_{\textbf{k}}(\tau) \excitor{k},
\end{equation}
and since all excitors commute, the nested commutator expansion truncates at $l=0$:
\begin{equation}
  \deriv{}{\tau}\expT{\hat{T}}
  =
  \expT{\hat{T}}
  \dot{\hat{T}}
  =
  \expT{\hat{T}}
  \left\lbrace \sum_{\textbf{j}} \dot{t}_{\textbf{j}}\excitor{j} \right\rbrace.
\end{equation}
The imaginary-time Schr\"{o}dinger equation \eqref{eq:cc-imaginary-time-schro}
then becomes:
\begin{equation}
  \expT{\hat{T}}\dot{\hat{T}}\refd
  =
  -\hat{H} \expT{\hat{T}}\refd,
\end{equation}
and upon projection onto $\bra{\Det{k}}\expT{-\hat{T}(\tau)} = \bra{\Det{0}}\excitor{k}^\dagger\expT{-\hat{T}(\tau)}$:
\begin{equation}\label{eq:ode-cc}
  \dot{t}_{\textbf{k}} = -\braket{\Det{0} | \excitor{k}^\dagger\bar{H}(\tau) | \Det{0} } = - \omega_{\textbf{k}}(\tau),
\end{equation}
since by construction $\braket{\Det{0} | \excitor{k}^\dagger \excitor{j} |
\Det{0}} = \delta_{\textbf{kj}}$. Equation \eqref{eq:ode-cc} is an
imaginary-time \ac{ODE} which we can solve by discretization.

The stochastic propagation of the linked \ac{CC} equations is thus directly
related to those utilised within \ac{FCIQMC},\cite{Booth2009}
\ac{DMC},\cite{Foulkes2001-lh, Toulouse2015-yz} and the original unlinked \ac{CCMC}
approach.\cite{Thom2010, Spencer2015a}
This allows us to understand limits on the time-step due to the
spectral range of the Hamiltonian, and more directly compare computational costs
with prior stochastic coupled cluster theory.

\subsection{Discretized imaginary-time propagation and preconditioning}
\label{sec:imag-time-and-precond}

The imaginary-time \ac{ODE} in Eq.~(\ref{eq:ode-cc}) can be discretized in a
number of ways. In principle, we would like to:
\begin{enumerate*}[label=\emph{\alph*})]
 \item use as large a time-step as possible without losing stability of the
  integrator, and
 \item perform the fewest possible number of evaluations of the \ac{CC} vector
   function per time-step. 
\end{enumerate*}
The usual approach in \ac{CC}\ac{MC} and \ac{FCIQMC} is the explicit Euler
method with a time-step $h$:
\begin{equation}\label{eq:explicit-Euler}
  t_{\textbf{k}}^{[n+1]} = t_{\textbf{k}}^{[n]} - h \omega_{\textbf{k}}^{[n]},
\end{equation}
where $t_{\textbf{k}}^{[n+1]}$ and $t_{\textbf{k}}^{[n]}$ are the cluster
amplitudes at times $\tau + h$ and $\tau$, respectively and
$\omega_{\textbf{k}}^{[n]}$ is the \ac{CC} vector function at time $\tau$.

Alternatively, one could use an implicit Euler scheme:
\begin{equation}\label{eq:implicit-Euler}
  t_{\textbf{k}}^{[n+1]} = t_{\textbf{k}}^{[n]} - h \omega_{\textbf{k}}^{[n+1]},
\end{equation}
where the right-hand side now depends on the \ac{CC} vector function evaluated
at time $\tau + h$. We can approximate this term using the Newtown--Raphson
step:\cite{Helgaker2000-yb}   
\begin{equation}
  \omega^{[n+1]}_{\textbf{k}} \simeq \omega^{[n]}_{\textbf{k}}
  + \sum_{\textbf{l}}A_{\textbf{kl}}^{[n]}\Delta t_{\textbf{l}}^{[n]} 
\end{equation}
where the \ac{CC} Jacobian has been introduced:
\begin{equation}\label{eq:jacobian}
    A_{\textbf{kl}}^{[n]} = 
 \braket{\Det{0} | \excitor{k}^{\dagger} \BCHfirst{\iterate{\bar{H}}{n}}{\excitor{l}} | \Det{0}}.
\end{equation}
and obtain the Rosenbrock--Euler method:\cite{Hairer1996-wl,Jarlebring2014-ke} 
\begin{equation}\label{eq:NR-implicit-Euler}
  t_{\textbf{k}}^{[n+1]} = t_{\textbf{k}}^{[n]}
  - h \sum_{\textbf{l}}\left[ \mat{I} + h\mat{A}^{[n]} \right]^{-1}_{\textbf{kl}} \omega_{\textbf{l}}^{[n]}.
\end{equation}
Under the assumption of non-singular Jacobian, we can use a Woodbury-type
identity to compute the inverse:\cite{Henderson1981-rj} 
\begin{equation}
  \left[ \mat{I} + h\mat{A}^{[n]} \right]^{-1} =  
  \left( h\mat{A}^{[n]} \right)^{-1} 
 -   
 \left( h\mat{A}^{[n]} \right)^{-1} 
 \left( h\mat{A}^{[n]} \right)^{-1} 
 \left[ \mat{I} + \left( h\mat{A}^{[n]} \right)^{-1} \right]^{-1}, 
\end{equation}
and retaining the first term only yields the \emph{deterministic}
Newtwon--Raphson step:\cite{Helgaker2000-yb}
\begin{equation}\label{eq:newton-step}
  t_{\textbf{k}}^{[n+1]} = t_{\textbf{k}}^{[n]}
  - \sum_{\textbf{l}}\left[ \mat{A}^{[n]} \right]^{-1}_{\textbf{kl}} \omega_{\textbf{l}}^{[n]}.
\end{equation}
Given this point of view, it is possible to relate the imaginary-time
propagation to a number of standard techniques in numerical analysis. Given a
time-step $\delta \tau$, the generalized step:
\begin{equation}\label{eq:relaxed-newton-step}
  t_{\textbf{k}}^{[n+1]} = t_{\textbf{k}}^{[n]}
  -
  \sum_{\textbf{l}}
  \delta \tau\left[ \mat{A}^{[n]} \right]^{-1}_{\textbf{kl}} \omega_{\textbf{l}}^{[n]},
\end{equation}
will be equivalent to a \emph{relaxed} Newton--Raphson method.

The use of the full \ac{CC} Jacobian for preconditioning would be extremely
expensive and a more pragmatic route is taken in practice.
The simplest choice is to approximate the Jacobian with the identity matrix,
\ie no preconditioning is applied to the iterations.
A more sophisticated approach is to only retain iteration-independent terms in
Eq.~\eqref{eq:jacobian}:
\begin{equation}\label{eq:zero-jacobian}
  A_{\textbf{kl}}^{[n]}
  \simeq
  A_{\textbf{kl}}
  =
  \delta_{\textbf{kl}}
  \braket{\Det{0} | \excitor{k}^{\dagger} \BCHfirst{\hat{H}^{\textrm{d}}}{\excitor{k}} | \Det{0}}
  +
  ( 1 - \delta_{\textbf{kl}} )
  \braket{\Det{0} | \excitor{k}^{\dagger} \BCHfirst{\hat{H}^{\textrm{od}}}{\excitor{l}} | \Det{0}},
\end{equation}
where the ``d'' and ``od'' stand for diagonal and off-diagonal, respectively.
We can then propose two cheap preconditioners. We can either use the diagonal
part of the Fock operator:\footnote{This is the zeroth-order Hamiltonian in a \ac{MP}
partitioning.}
\begin{equation}\label{eq:fock-preconditioner}
  A_{\textbf{kl}}^{\textrm{Fock}}
  \simeq
  \delta_{\textbf{kl}}
  \braket{\Det{0} | \excitor{k}^{\dagger} \BCHfirst{\hat{F}^{\textrm{d}}}{\excitor{k}} | \Det{0}}
  \leftarrow
  \left\lbrace
  \diagram{
    \draw (-0.5,0) node[circlex] (f) {} -- (0,0) node[dot=white] (f1) {};
    \draw[->-] (f1) to node[right] {} ++(0,+0.5);
    \draw[-<-] (f1) to node[right] {} ++(0,-0.5);
  },
  \diagram{
    \draw (-0.5,0) node[circlex] (f) {} -- (0,0) node[dot=white] (f1) {};
    \draw[-<-] (f1) to node[right] {} ++(0,+0.5);
    \draw[->-] (f1) to node[right] {} ++(0,-0.5);
  }
  \right\rbrace,
\end{equation}
or the diagonal part of the full Hamiltonian:
\begin{equation}\label{eq:full-preconditioner}
  A_{\textbf{kl}}^{\textrm{full}}
  \simeq
  \delta_{\textbf{kl}}
  \braket{\Det{0} | \excitor{k}^{\dagger} \BCHfirst{\hat{F}^{\textrm{d}} + \hat{\Phi}^{\textrm{d}}}{\excitor{k}} | \Det{0}}
  \leftarrow
  \left\lbrace
  \diagram{
    \draw (-0.5,0) node[circlex] (f) {} -- (0,0) node[dot=white] (f1) {};
    \draw[->-] (f1) to node[right] {} ++(0,+0.5);
    \draw[-<-] (f1) to node[right] {} ++(0,-0.5);
  },
  \diagram{
    \draw (-0.5,0) node[circlex] (f) {} -- (0,0) node[dot=white] (f1) {};
    \draw[-<-] (f1) to node[right] {} ++(0,+0.5);
    \draw[->-] (f1) to node[right] {} ++(0,-0.5);
  },
  \diagram{
    \interaction{2}{g}{(0,0)}{dot=white}{flexdotted};
    \draw[->-] (g1) to node[left] {} ++(0,+0.5);
    \draw[-<-] (g1) to node[left] {}++(0,-0.5);
    \draw[->-] (g2) to node[left] {}++(0,+0.5);
    \draw[-<-] (g2) to node[left] {}++(0,-0.5);
  },
  \diagram{
    \interaction{2}{g}{(0,0)}{dot=white}{flexdotted};
    \draw[-<-] (g1) to node[left] {} ++(0,+0.5);
    \draw[->-] (g1) to node[left] {}++(0,-0.5);
    \draw[-<-] (g2) to node[left] {}++(0,+0.5);
    \draw[->-] (g2) to node[left] {}++(0,-0.5);
  },
  \diagram{
    \interaction{2}{g}{(0,0)}{dot=white}{flexdotted};
    \draw[->-] (g1) to node[left] {} ++(0,+0.5);
    \draw[-<-] (g1) to node[left] {}++(0,-0.5);
    \draw[-<-] (g2) to node[left] {}++(0,+0.5);
    \draw[->-] (g2) to node[left] {}++(0,-0.5);
  }
  \right\rbrace.
\end{equation}
The former is universally implemented in deterministic \ac{CC} codes and its
effectiveness can be justified through perturbative
arguments.\cite{Helgaker2000-yb} Use of the latter has not, to the best of our
knowledge, been attempted before.

The derivation here presented makes explicit the connection with preconditioning
already discussed by some of us in connection with
\ac{FCIQMC}\cite{Blunt2019-tf} and unlinked \ac{CCMC}.\cite{Neufeld2020-og} We
will discuss how preconditioning is implemented for diagCCMC in Section
\ref{sec:precond}.

Finally, let us point out that Jarlebring \emph{et al.} showed how a specific
instance of a nonlinear eigenvalue problem is equivalent to a Rosenbrock-type
discretization of an associated imaginary-time \ac{ODE}.\cite{Jarlebring2014-ke}
An adaptive time-step integrator can be thus formulated based on convergence
estimates similar to those presented in ref.~\citenum{Jarlebring2014-ke}. 

\section{Diagrammatic Coupled Cluster Monte Carlo}

We wish to stochastically solve the linked \ac{CC}
equations~\eqref{eq:linked-cc-amplitudes}. Additionally, and at variance with the approach of Franklin
\emph{et al.}, we wish to overcome the need for a corrected update step and
the sampling of extraneous unlinked terms.\cite{Franklin2016}
Whereas the latter have been observed to cancel out on average, they impose
limitations to what system sizes are approachable before the memory cost
becomes prohibitive.

In the diagCCMC algorithm\cite{Scott2019-ge} we use the \emph{uncorrected}
update step in Eq.~(\ref{eq:explicit-Euler}). Two novel insights allow us to achieve this goal:
\begin{itemize}
  \item The \ac{CC} wavefunction is stored in a compressed representation
    \emph{without} invoking particles or walkers. It is comparatively easier to
    enforce constant unit intermediate normalization within a walker-less
    algorithm.
  \item The \ac{CC} vector function appearing in the update step is an integral
    expressible as a terminating series expansion. Terms in this expansion can
    be evaluated stochastically. 
\end{itemize}
The use of diagrammatic techniques automatically guarantees that only connected
terms in the similarity-transformed Hamiltonian are included.
The sampling will thus happen in ``diagram space'' and relies on the even
selection algorithm of Scott \emph{et al.} \cite{Scott2017}

\subsection{Stochastic compression without walkers}\label{sec:walker-less}

Previous algorithms to stochastically solve the linked \ac{CC} equations
modified the propagation in~\eqref{eq:explicit-Euler} to approach the correct
solution. The need for such modifications can be attributed to the use of a
\emph{variable} intermediate normalization:
\begin{equation}\label{eq:ccmc-prime-wave}
  \ket{\text{CCMC}} = N_{0}\expT{\frac{\hat{\tilde{T}}}{N_{0}}} \refd,
\end{equation}
where the additional normalization parameter $N_{0}$ is constrained by the energy
equation:
\begin{equation}
 N_{0}\braket{D_{\textbf{0}} | \bar{H} - E_{\mathrm{CC}} | D_{\textbf{0}}}  = 0, 
\end{equation} 
and the unknown \ac{CC} energy has to be substituted by the shift $S$.
At the beginning of the simulation, $S = E_{\mathrm{ref}}$ and this causes the energy estimator
to converge incorrectly prior to initialisation of population control.
However, upon closer inspection, the wavefunction Ansatz in
\eqref{eq:ccmc-prime-wave} is seen to be equivalent to the conventional \ac{CC}
Ansatz with:
\begin{enumerate*}[label=\emph{\alph*})]
  \item overlap with the reference set to $N_{0}$, and
  \item all nonzero cluster amplitudes $t_{\textbf{k}}$ represented by values
    larger than $\frac{1}{N_{0}}$.
\end{enumerate*}
The floating intermediate normalization can then be interpreted as an algorithmic
choice to determine the \emph{granularity} of representation during the
calculation and achieve compression of the \ac{CC} wavefunction.
This choice is arbitrary and can be related back to the conventional \ac{CC}
Ansatz.
Assume then that the intermediate normalization is now a constant value
$\braket{D_{\mathbf{0}}| \text{CCMC}} = N_{0}$, set as an input parameter to the
calculation. At sufficiently small granularities, the calculation will
spontaneously stabilise at a system-dependent population of walkers, without
the need for population control. The stochastic realization of the modified explicit Euler integration:
\begin{equation}
  \tilde{t}_{\textbf{k}}^{[n+1]} = \tilde{t}_{\textbf{k}}^{[n]} - \delta \tau N_{0} \tilde{\omega}_{\textbf{k}}^{[n]}
\end{equation}
would then take the form:
\begin{enumerate}
  \item \textbf{Compress} the cluster amplitudes to the selected granularity, by
    \emph{stochastically} rounding those
    amplitudes for which $|\tilde{t}_{\mathbf{k}}| < 1$ to $\sgn(\tilde{t}_{\mathbf{k}})\times
    1$ or $0$, 
  \item \textbf{Evaluate} the \ac{CC} vector function by taking a large enough number of
    samples such that diagrams in $\tilde{\omega}_{\textbf{k}}$ of magnitude $1$
    are, on average, selected once.
  \item \textbf{Adjust} the time-step $\delta \tau$ as to avoid particle blooms, that is large
spawning events which would destabilise the calculation dynamics. 
\end{enumerate}
We can however take one further step and cast away the walker interpretation
entirely. The thresholding implied in the previous algorithmic sketch can be
rigorously formulated \emph{without} recourse to walkers. 
We introduce three strictly positive calculation parameters: the
\emph{representation} granularity, $\Delta$, the \emph{evaluation} granularity,
$\gamma$, and the maximum diagram contribution $\epsilon$. The algorithm then will:
\begin{enumerate}
\item \textbf{Compress} the cluster amplitudes to the chosen
  representation granularity, by stochastic rounding amplitudes for which
  $|\tampc{k}| < \Delta$ to $\sgn(\tampc{k})\times \Delta$ or $0$.
\item \textbf{Evaluate} the \ac{CC} vector function stochastically such that diagrams
  with magnitude $\gamma$ are selected once on average. 
\item \textbf{Adjust} the time-step $\delta\tau$ such that the maximum diagram contribution,
  $\delta\tau\frac{w_{\mathrm{diagram}}}{p_{\mathrm{diagram}}}$, is of magnitude $\epsilon$.
\end{enumerate}
The walker and walker-less representations are entirely equivalent. The
representation granularity is the inverse of the intermediate normalization
constant $\Delta = \frac{1}{N_0}$, the condition $\gamma = \Delta$ defines the
even selection approach \cite{Scott2017}, and the ratio
$\frac{\epsilon}{\Delta}$ is the maximum allowed size for a spawning event.
The resultant approach to the imposition of sparsity bears some resemblance to recent Fast Randomized Iteration approaches.\citep{Lim2017fast,Greene2019-ou}

Within this approach the total walker population is the sum of rescaled cluster
coefficient absolute magnitudes and the reference
$\frac{1}{\Delta}+\sum_{\textbf{i}}\frac{|t_{\textbf{i}}|}{\Delta}$. It is thus
not needed to set the hard-to-predict total walker population as a calculation
parameter: choosing to stochastically round all cluster coefficients below a
certain value gives a more intuitively stable treatment between different
calculations.
The total walker population can vary dramatically with system size:
evaluating and comparing computational cost and performance for systems of
varying size can be a nontrivial challenge. Instead we expect the walker-less picture
to manifest the \emph{transferability} property of cluster amplitudes:
\cite{Flocke2004-qm} the magnitude of the amplitudes should be relatively
unchanged with system size, especially when localised orbitals are used, allowing
equivalent parameters for different calculations to be easily identified.

We have found $\gamma = 10^{-3}$ to be the lowest evaluation granularity giving
a calculation stable enough to extract statistics from. While smaller $\gamma$ values
achieve more stable calculations, with $10^{-4}$ providing a reasonable
compromise between computational cost and stability.
We have also continued to use conventions from the particle representation for
now by setting $\frac{\Delta}{\gamma} = 1$ and $\frac{\epsilon}{\gamma} = 3$.

\subsection{Selection of diagrams}\label{sec:diagrammatic-sampling}

The second essential insight enabling the diagCCMC algorithm is the stochastic
evaluation of the \ac{CC} vector function on the right-hand side of the
uncorrected update step.
At any given excitation level in the \ac{CC} hierarchy, the \ac{BCH} expansion
of $\bar{H}$ will truncate at the four-fold nested commutator:
$\omega_{\textbf{k}}$ is expressible as a sum of a finite, enumerable number of
terms. We choose to represent these terms as diagrams and generate such an
expansion on-the-fly, rather than enumerating the allowed diagrams beforehand.
In each main Monte Carlo cycle in the algorithm, we perform the evaluation of
the integral by attempting to select $n_{\mathrm{a}}$ \emph{fully specified}
diagrams from its expansion. 
The action of the similarity-transformed Hamiltonian on the reference
determinant can be written compactly as:
\begin{equation}
  \bar{H}\refd = w_{\textbf{l}}\excitor{l}\refd  
\end{equation}
where the amplitude $w_{\textbf{l}} = w_{\mathrm{H}}\prod_{m} \tampc{m}$ is a
product of a one- or two-body integral from the Hamiltonian and a cluster of
excitors. The multi-index $\textbf{l}$ is fully specified, meaning that all
hole and particle lines are \emph{explicitly} labelled. 
The amplitude is determined by the contraction pattern randomly
selected during diagram generation. Finally, since 
$\braket{\Det{0} | \excitor{p}^{\dagger}\excitor{q}| \Det{0}} =
\delta_{\textbf{pq}}$, the selected diagram can contribute to one and only one
cluster amplitude: the one whose multi-index corresponds to the \emph{external}
lines in $w_{\textbf{l}}$.
The rules for the deterministic enumeration of diagrams that were briefly
detailed in Section \ref{sec:diagrammatic-representation} are largely
unmodified in our stochastic algorithm. Each step corresponds to an event
occurring with an easily computed probability:
\begin{enumerate}
\item Sample the action of the wave operator: with probability
  $p_{\mathrm{sel}}$, choose a
  term from the \ac{BCH} expansion \eqref{eq:H-commutator-expansion}, that is
  select a cluster of size $N \leq 4$ and the excitation level of each
  constituent excitor. We use the even selection scheme of Scott \emph{et al.}
  \cite{Scott2017} in a walker-less representation, see Section~\ref{sec:walker-less-even}.  
\item Sample the action of the Hamiltonian: with probability
  $p_{\mathrm{hver}}$ choose one of the 13 interaction vertices in
  Table~\ref{tab:H-vertices}. This step is not independent of the former and we
  use importance sampling, see Section~\ref{sec:walker-less-even}.
\item Sample the admissible contraction patterns: with probability
  $p_{\mathrm{cont}}$, choose a specific Kucharski--Bartlett sign sequence
  \cite{Kucharski1986-sz, Crawford2000, Shavitt2009-mr}, see Table
  \ref{tab:Phi-T12-T3} for an example. 
\item Sample the index set to label internal lines. Given the number of internal hole and
  particle lines in the selected contraction, the probability associated to this
  step $p_{\mathrm{int}}$ is computed combinatorically.
\item Sample the index set to label external lines. As for the previous step,  
  the probability $p_{\mathrm{ext}}$ is also computed combinatorically.
\end{enumerate}
With this process, we are able to obtain a given diagram with probability
$p_{\mathrm{diagram}} =
p_{\mathrm{select}}p_{\mathrm{hver}}p_{\mathrm{cont}}p_{\mathrm{int}}p_{\mathrm{ext}}$
and in each Monte Carlo step the diagram is sampled $p_{\mathrm{diagram}}\times
n_a$ times.

\begin{table}
 \caption{Generation of diagrams 
   stemming from the $(\Phi T_1^2T_3)_\mathrm{c}$ term and contributing to the
   triples equations. The Kucharski--Bartlett sign sequences for excitor and
   interaction vertices, resulting contraction patterns, and resulting diagrams
   are shown.}\label{tab:Phi-T12-T3}
 \centering
 \begin{ruledtabular}
   \begin{tabular}{c c c c}
     Excitors & Interaction & Contraction & Diagram \\
     \multirow{5}{*}{$+-|+-|+++---$} & \multirow{5}{*}{$++--$} & $+-|+|-$ &
$
\diagram{
  \interaction{1}{ta}{(0,-0.25)}{ddot}{overhang};
  \interaction{1}{tb}{(1,-0.25)}{ddot}{overhang};
  \interaction{3}{tc}{(2,-0.25)}{ddot}{overhang};
  \draw[flexdotted] (0, 0.25) node[dot=white] (g1) {} to (1.5,0.25) node[dot=white] (g2) {};
  \draw[semithick, bend left] (ta1) to (g1); 
  \draw[semithick, bend right] (ta1) to (g1); 
  \draw[semithick] (tb1) to ++(-0.25,0.75);
  \draw[->-] (tb1) to (g2);
  \draw[->-] (g2) to (tc1);
  \draw[semithick] (tc1) to ++(0.25,0.75);  
  \draw[semithick] (tc2) to ++(-0.25,0.75);  
  \draw[semithick] (tc2) to ++(0.25,0.75);  
  \draw[semithick] (tc3) to ++(-0.25,0.75);  
  \draw[semithick] (tc3) to ++(0.25,0.75);  
}
$ \\[.2cm]
                                   &                         & $+-|-|+$ &
$
\diagram{
  \interaction{1}{ta}{(0,-0.25)}{ddot}{overhang};
  \interaction{1}{tb}{(1,-0.25)}{ddot}{overhang};
  \interaction{3}{tc}{(2,-0.25)}{ddot}{overhang};
  \draw[flexdotted] (0, 0.25) node[dot=white] (g1) {} to (1.5,0.25) node[dot=white] (g2) {};
  \draw[semithick, bend left] (ta1) to (g1); 
  \draw[semithick, bend right] (ta1) to (g1); 
  \draw[semithick] (tb1) to ++(-0.25,0.75);
  \draw[-<-] (tb1) to (g2);
  \draw[-<-] (g2) to (tc1);
  \draw[semithick] (tc1) to ++(0.25,0.75);  
  \draw[semithick] (tc2) to ++(-0.25,0.75);  
  \draw[semithick] (tc2) to ++(0.25,0.75);  
  \draw[semithick] (tc3) to ++(-0.25,0.75);  
  \draw[semithick] (tc3) to ++(0.25,0.75);  
}
$ \\[.2cm]
                                   &                         & $+|--|+$ &
$
\diagram{
  \interaction{1}{ta}{(0,-0.25)}{ddot}{overhang};
  \interaction{3}{tb}{(1,-0.25)}{ddot}{overhang};
  \interaction{1}{tc}{(4,-0.25)}{ddot}{overhang};
  \draw[flexdotted] (0.5, 0.25) node[dot=white] (g1) {} to (3.5,0.25) node[dot=white] (g2) {};
  \draw[semithick] (ta1) to ++(-0.25,0.75);
  \draw[->-] (ta1) to (g1);                                                   
  \draw[semithick] (tb1) to ++(0.25,0.75);
  \draw[-<-] (tb1) to (g1);                                                   
  \draw[semithick] (tb2) to ++(0.25,0.75);  
  \draw[semithick] (tb2) to ++(-0.25,0.75);  
  \draw[semithick] (tb3) to ++(-0.25,0.75);  
  \draw[->-] (tb3) to (g2);                                                   
  \draw[semithick] (tc1) to ++(0.25,0.75);  
  \draw[-<-] (tc1) to (g2);                                                   
}
$ \\[.2cm]
                                   &                         & $-|++|-$ & 
$
\diagram{
  \interaction{1}{ta}{(0,-0.25)}{ddot}{overhang};
  \interaction{3}{tb}{(1,-0.25)}{ddot}{overhang};
  \interaction{1}{tc}{(4,-0.25)}{ddot}{overhang};
  \draw[flexdotted] (0.5, 0.25) node[dot=white] (g1) {} to (3.5,0.25) node[dot=white] (g2) {};
  \draw[semithick] (ta1) to ++(-0.25,0.75);
  \draw[-<-] (ta1) to (g1);                                                   
  \draw[semithick] (tb1) to ++(0.25,0.75);
  \draw[->-] (tb1) to (g1);                                                   
  \draw[semithick] (tb2) to ++(0.25,0.75);  
  \draw[semithick] (tb2) to ++(-0.25,0.75);  
  \draw[semithick] (tb3) to ++(-0.25,0.75);  
  \draw[-<-] (tb3) to (g2);                                                   
  \draw[semithick] (tc1) to ++(0.25,0.75);  
  \draw[->-] (tc1) to (g2);                                                   
}
$ \\[.2cm]
                                   &                         & $+|-|+-$ &
$
\diagram{
  \interaction{1}{ta}{(0,-0.25)}{ddot}{overhang};
  \interaction{1}{tb}{(1,-0.25)}{ddot}{overhang};
  \interaction{3}{tc}{(2,-0.25)}{ddot}{overhang};
  \draw[flexdotted] (0.5, 0.25) node[dot=white] (g1) {} to (2,0.25) node[dot=white] (g2) {};
  \draw[semithick] (ta1) to ++(-0.25,0.75);
  \draw[semithick] (ta1) to (g1);                                                   
  \draw[semithick] (tb1) to ++(0.25,0.75);
  \draw[semithick] (tb1) to (g1);                                                   
  \draw[semithick, bend left] (tc1) to (g2);  
  \draw[semithick, bend right] (tc1) to (g2);  
  \draw[semithick] (tc2) to ++(0.25,0.75);  
  \draw[semithick] (tc2) to ++(-0.25,0.75);  
  \draw[semithick] (tc3) to ++(0.25,0.75);  
  \draw[semithick] (tc3) to ++(-0.25,0.75);  
}
$ \\
   \end{tabular}
   \end{ruledtabular}
\end{table}

We need further minor modifications to the deterministic enumeration of diagrams
to ensure that the \ac{CC} vector function is evaluated correctly.
Our algorithm singles out specific diagrams, where all lines,
internal and external, are explicitly labelled.
This procedure identifies a single cluster amplitude $\tampc{k}$ to which the selected
diagram will contribute without having to sum over internal lines.
Permutational symmetries will thus have to be handled differently, such that the
our probability distributions are properly normalised. Sums 
of the form $\frac{1}{2}\sum\limits_{i,j}$ have to be replaced with 
$\sum\limits_{i>j} + \frac{1}{2}\delta_{ij}$ to ensure that there is only a
single way to select diagrams related by:
\begin{itemize}
\item the antipermutation of indices stemming from antisymmetrized interaction vertices.
\item the antipermutation of hole or particle indices stemming from excitor vertices.
\item the commutation of excitors.
\end{itemize}
For the first two cases, terms with $i=j$ would vanish
when summing over equivalent indices. In the last case, the diagonal case $i=j$
indicates additional symmetries of the resulting diagram. 
In our stochastic diagram enumeration, each pair of equivalent internal or
external lines will not require a $\frac{1}{2}$ factor. Moreover, upon
selection a well-defined ordering of excitors is established, which removes the
need for $\frac{1}{2}$ factors in diagrams where excitors of the same rank appear.
These modification to the deterministic evaluation rules ensure a unique
selection of a contraction pattern\cite{Kucharski1986-sz, Crawford2000,
  Shavitt2009-mr}. The action of permutation operators for inequivalent external
lines is subsumed into the permutation of hole and particle indices needed to
store the result of the sampling in antisymmetrised ordering, which provides the
appropriate parity factor $(-1)^{\sigma}$.
With these considerations, a contribution to $\tampc{k}$ is computed as:
\begin{equation}\label{eq:diagrammatic-update-contribution}
x_{\mathrm{diagram}} 
= \frac{w_{\mathrm{diagram}}}{p_{\mathrm{diagram}}} 
= \frac{(-1)^{\sigma}w_{\mathrm{T}}w_{\mathrm{H}}}{p_{\mathrm{select}}p_{\mathrm{hver}}p_{\mathrm{cont}}p_{\mathrm{int inds}}p_{\mathrm{ext inds}}}
\end{equation}
and the sampling algorithm is designed such that $p_{\mathrm{diagram}} =
|w_{\mathrm{diagram}}|$. Ultimately, our aim is to achieve importance sampling between contributions, see Section \ref{sec:importance-sampling}.

\subsection{Preconditioning}
\label{sec:precond}

While the imaginary-time propagation discussed in
Section~\ref{sec:stochastic-cc} will generally be used within our work, we could
also make use of arbitrary preconditioners. Apart from the identity, we
implemented two additional options: the diagonal of the Fock operator (diagrams 1 and 2 in Table \ref{tab:H-vertices}) and
the diagonal of the full Hamiltonian (diagrams 1, 2, 4, 5, and 6 in Table \ref{tab:H-vertices}).
The connected portions of these vertices do not modify excitors when applied.
These preconditioners are the iteration-independent
approximations to the Jacobian discussed in Section~\ref{sec:imag-time-and-precond} and strike
a balance between computational complexity and improvement of convergence. The
diagonal Fock preconditioner is ubiquitously implemented in deterministic
\ac{CC} approaches.\cite{Helgaker2000-yb}
Since all relevant quantities are precomputed the values of either
preconditioner can be evaluated with a cost independent of system size, unlike
implementation of the similar approach within \ac{FCIQMC} and
\ac{CCMC}.\citep{Blunt2019-tf,Neufeld2020-og}

The portion of the Hamiltonian used for preconditioning can then be applied via
a straightforward rescaling of the original cluster amplitudes by a factor of $1 -\delta \tau$.
The remainder of the Hamiltonian is applied explicitly, as in imaginary time
propagation, before rescaling by the preconditioner.

We will not investigate the benefits of preconditioning here, but wanted to observe that
the diagrammatic formalism lends itself to a straightforward implementation of a range 
of preconditioners without introducing additional computational costs scaling with
system size. This results from the use of the connected portions of all preconditioners,
unlike previous stochastic approaches.\citep{Blunt2019-tf,Neufeld2020-og}

\section{$\bar{H}$, importance sampling, and even selection}

The even selection algorithm was proposed by Scott and Thom \cite{Scott2017} to
improve the sampling of the action of the \ac{CC} wave operator on the reference
determinant. Even selection was specifically designed to alleviate calculation
instabilities due to the occurrence of large particle blooms.
Sampling proceeds \emph{via} selection of clusters containing a specific number of excitors of each
rank, termed a \emph{combination}, separately. 
In this section, we summarise the adaptation of even selection in a 
walker-less context. We then illustrate the need for importance sampling of
$(H\expT{T})_{\mathrm{c}}$ and describe the strategy implemented in diagCCMC.

\subsection{Walker-less Even Selection}\label{sec:walker-less-even}

The original even selection algorithm defined the probability of selecting a
particular set of excitors $e$ from combination $c$ of size $s$ as:
\begin{equation}
  p_{\textrm{select}}(e) 
  =
  p_{\textrm{size}}(s)p_{\textrm{combo}}(c|s)p_{\textrm{excitors}}(e|c,s),
\end{equation}
a series of conditional probabilities. 
In the following, we re-express the selection probability as:
\begin{equation}
  p_{\textrm{select}}(e) = p_{\textrm{combo}}(c)p_{\textrm{excitors}}(e|c),
\end{equation}
to simplify considerations to follow. We also assume unit intermediate normalization.

We adopt the same notation used in ref.~\citenum{Scott2017} and denote the number
of excitors of rank $j$ within combination $c$ as $\eta_{cj}$.
$L_{j}$ is the sum of absolute magnitudes of cluster amplitudes at rank $j$,
that is, the $\ell_{1}$-norm of $T_{j}$.

In keeping with the original approach, $\eta_{cj}$ denotes the number of
excitors of rank $j$ contained within combination $c$, $L_j$ the sum of cluster
coefficient absolute magnitudes at rank $j$, and $n_{\mathrm{a}}$ the number of sampling
attempts to be made that iteration. 

In the walker-less representation the amplitude of a given cluster is the
product of cluster coefficients $w_{e} = \prod_{\textbf{p}}\tampc{p}$.
The evaluation granularity is, by definition, equal to the absolute magnitude
of the \ac{MC} weight: 
\begin{equation}\label{eq:stoch-gran-definition}
|x_e| = \frac{|w_e|}{n_{\textrm{a}}p_{\textrm{select}}(e)} = \gamma.
\end{equation}
Even selection for all clusters requires that evaluation and representation
granularities be the same: $\gamma = \Delta$.
As we also require $\frac{|w_e|}{p_{\textrm{select}}(e)}$ to be an
excitor-independent constant we obtain:
\begin{align}
p_{\textrm{excitors}}(e|c) &=  \prod\limits_{j=1}^{l} \left( \eta_{cj}!\prod\limits_{\textbf{p}\in e_{j}}^{\eta_{cj}} \frac{|\tampc{p}|}{L_{j}}\right)\\
p_{\textrm{combo}}(c) &= \frac{1}{W} \prod\limits_{j=1}^{l}  \frac{L_{j}^{\eta_{cj}}}{\eta_{cj}!}
\end{align}                        
with normalization constant:
\begin{equation*}
 W = \sum\limits_c^{n_{\textrm{combo}}} \prod\limits_{j=1}^{l}  \frac{L_{j}^{\eta_{cj}}}{\eta_{cj}!}
\end{equation*}
and where $n_{\textrm{combo}}$ is the total number of combinations.
The number of random samples to take within a calculation is thus obtained from
the evaluation granularity given as input:
\begin{equation}\label{eq:nattempts-definition}
n_{\textrm{a}} = \frac{W}{\gamma}.
\end{equation}

\subsection{Motivation for importance sampling}
\label{subsec:importance_sampling_motiv}
Each pairing of excitor combinations with Hamiltonian vertices can result in a
different number of admissible contractions and thus fully indexed diagrams. It
it non-trivial to ensure that $|x_{\mathrm{diagram}}|$ in
\eqref{eq:diagrammatic-update-contribution} is in any sense comparable between
the different pairings.
The selection of a Hamiltonian interaction vertex is not independent of the
selection of excitor combination: $p_{\mathrm{hver}}$ will be a probability
conditional on $p_{\textrm{select}}$.
This enables the use of \emph{truncated excitation generation}, that is the \emph{a
priori} exclusion of Hamiltonian-excitor pairings which will not be able to
contribute to any stored amplitude\cite{Neufeld2019-sb}. Furthermore, one could
easily exclude any class of diagrams that we wish to evaluate with a different algorithm.

Clusters from different combinations can contribute to a set of allowed
diagrams, whose number can undergo large variations, especially with varying
system size.
In a \ac{CC} calculation to any order let us consider two limiting cases
in the sampling to clarify this statement.
Assume that we selected a cluster from the combination $T_{1}^{4}$ with no
repeated excitors. The only fully connected diagrams stemming from such a
cluster are of the form:
\begin{equation}
  \diagram{
    \interaction{1}{ta}{(0, -0.25)}{ddot}{overhang};
    \interaction{1}{tb}{(1, -0.25)}{ddot}{overhang};
    \interaction{1}{tc}{(2, -0.25)}{ddot}{overhang};
    \interaction{1}{td}{(3, -0.25)}{ddot}{overhang};
    \draw[flexdotted] (0.5, 0.25) node[dot=white] (g1) {} to (2.5, 0.25) node[dot=white] (g2) {};
    \draw[semithick] (ta1) to ++(-0.25, 0.75);
    \draw[semithick] (ta1) to (g1);                                                   
    \draw[semithick] (g1) to (tb1);                                                   
    \draw[semithick] (tb1) to ++(0.25, 0.75);
    \draw[semithick] (tc1) to ++(-0.25, 0.75);  
    \draw[semithick] (tc1) to (g2);  
    \draw[semithick] (g2) to (td1);                                                   
    \draw[semithick] (td1) to ++(0.25, 0.75);
  }
\end{equation}
as such: 
\begin{enumerate}
\item only one Hamiltonian interaction vertex is admissible: 
$\diagram{
    \interaction{2}{g}{(0,0)}{dot=white}{flexdotted};
    \draw (g1) to node[left] {} ++(-0.25,-0.5);
    \draw (g1) to node[right] {} ++(0.25,-0.5);
    \draw (g2) to node[left] {} ++(-0.25,-0.5);
    \draw (g2) to node[right] {} ++(0.25,-0.5);
  }$ with probability $p_{\textrm{hver}} = 1.0$
\item selecting a contraction pattern boils down to the choice of which two excitors from
  the four to be connected to the interaction vertex via hole-type lines. 
  There are $\binom{4}{2}=6$ possible ways of doing so, which gives: $p_{\textrm{cont}} = \frac{1}{6}$.
\item being single excitations, each excitor has one particle and one hole
  line. Once the contraction pattern is set there is only one choice to make per
  hole line in the diagram and each will be made with probabililty
  $p_{\textrm{int}} = 1.0$.
\item As all external indices are fully determined by the selected cluster and contraction, there is only a single possible choice of external indices, so $p_{\textrm{ext}} = 1.0$.
\end{enumerate}
Each cluster from a $T_{1}^{4}$ combination can contribute to 6 valid diagrams,
independently of system size and truncation level. If all diagrams are selected
without weighting: $p_{\textrm{diagram}} = \frac{1}{6}p_{\textrm{select}}$. 

Now assume instead that we are sampling the action of a bare Hamiltonian vertex,
that is a cluster of size 0. There are only two admissible choices in such a
case:
\begin{alignat}{2}
  \diagram{
    \interaction{2}{g}{(0,0)}{dot=white}{flexdotted};
    \draw[-<-] (g1) to node[left] {} ++(-0.25,0.5);
    \draw[->-] (g1) to node[right] {} ++(0.25,0.5);
    \draw[-<-] (g2) to node[left] {} ++(-0.25,0.5);
    \draw[->-] (g2) to node[right] {} ++(0.25,0.5);
  }\quad
  \text{or}
  &\quad
  \diagram{
    \draw (-0.5,0) node[circlex] (f) {} -- (0,0) node[dot=white] (f1) {};
    \draw[-<-] (f1) to node[left] {} ++(-0.25,0.5);
    \draw[->-] (f1) to node[right] {} ++(0.25,0.5);
  }
\end{alignat}
There is no contraction to decide upon and hence no internal indices to decide
upon: $p_{\textrm{cont}} = 1.0 = p_{\textrm{int}}$. 
However, the amount of such terms to sample varies with system size. For an $N$-electron system with
$V$ virtual orbitals, there are $O(N)$ possible external hole and $O(V)$
possible external particle indices, respectively. 
For a one-body interaction vertex, there are $O(NV)$ admissible labelings of the
diagram, while for the two-body case there are $O(N^{2}V^{2})$ such labelings.

These examples show how sampling different classes of diagrams will require a varying
number of attempts $n_{\textrm{a}}$ in each \ac{MC} step. The original 
even selection prescription will need to be modified to accommodate this, or else calculations
will rapidly become untenably expensive.

\subsection{Importance sampling of clusters and Hamiltonian vertices}
\label{sec:importance-sampling}

To compensate for the difference in diagram generation between different
combinations we will now modify our sampling to include combination-dependent
constants $\alpha_c$:
\begin{alignat}{2}
  0.0 < \alpha_c \leq 1.0, \quad \max(\alpha_c) = 1.0
\end{alignat}
 such that:
\begin{alignat}{2} 
  p_{\textrm{combo}}(c) 
= 
\frac{\alpha_c}{W} \prod\limits_{j=1}^{l}\frac{L_{j}^{\eta_{cj}}}{\eta_{cj}!}, 
\quad
W = \sum\limits_c^{n_{\textrm{combo}}} \alpha_c\prod\limits_{j=1}^{l}  \frac{L_{j}^{\eta_{cj}}}{\eta_{cj}!}
\end{alignat}   
Each combination will be evaluated to a different granularity $\gamma_{c}$
defined as in \eqref{eq:stoch-gran-definition}:
\begin{equation}
\gamma_c 
= 
\left.\frac{|w_e|}{n_ap_{\textrm{select}}(e)}\right|_{e\in c} =\frac{W}{n_a\alpha_c}
= 
\frac{\gamma}{\alpha_c}.
\end{equation}
where we used the value of $n_{\textrm{a}}$ given in \eqref{eq:nattempts-definition}.

With this modification, we now require an approximately constant contribution to
the integrals $\langle D_0 |\excitor{k}^\dagger \bar{H}|D_{0}\rangle$. From section
\ref{sec:diagrammatic-sampling} we know the diagram amplitude:
\begin{equation}
|x_{\textrm{diagram}}|
=
\frac{|w_{\textrm{e}}|}{n_ap_{\textrm{select}}(e)}\frac{|w_{\textrm{H}}|}{p_{\textrm{hver}}p_{\textrm{cont}}p_{\textrm{int}}p_{\textrm{ext}}},
\end{equation}
must then be a constant $\Xi$.
This constant is a product of interaction vertex-specific and 
diagram-specific factors:
\begin{equation}
\Xi = \frac{\gamma_c}{\zeta_{ci}} B_{ci},
\end{equation}
where $\zeta_{ci}$ is the probability of choosing the $i$-th Hamiltonian interaction
vertex (table \ref{tab:H-vertices}) when sampling excitor combination $c$ 
and 
\begin{equation}
  B_{ci}= \frac{|w_{\textrm{H}}|}{p_{\textrm{cont}}p_{\textrm{int}}p_{\textrm{ext}}}.
\end{equation}
The $\zeta_{ci}$ probabilities are normalized: $\sum_{i} \zeta_{ci} = 1$.

Unfortunately, having \emph{all} contributions be of constant
magnitude is not a tenable aim. We can instead aim to
have either the average or maximum contribution from the sampling of each
combination with each admissible interaction vertex be a constant value. 
We thus require:
\begin{equation}
  \Xi = \frac{\gamma_c}{\zeta_{ci}} B_{ci}^{\star} = \frac{\gamma}{\alpha_c\zeta_{ci}} B_{ci}^{\star}
\end{equation}
where $B_{ci}^{\star}$ can be either the maximum or the average value for the
diagram-specific term.
Considering the maximum contribution, we then obtain:
\begin{subequations}
  \begin{align}
    \gamma_{ci} &= \frac{B_{ci}^{\textrm{max}}}{\sum_jB_{cj}^{\textrm{max}}}\\
    \chi_c &= \frac{B_{ci}^{\textrm{max}}}{\zeta_{ci}}\\
    \alpha_c &= \frac{\chi_c}{\max(\chi_c)}\\
    \Xi &= \frac{\gamma}{\max(\chi_c)},
  \end{align}
\end{subequations}
and similarly for the average contribution involving $B_{ci}^{\textrm{ave}}$.
Utilising these expressions requires on-the-fly accumulation of the values
$B_{ci}^{\star}$ during a calculation.
$\Xi$ will the be the value of the maximum (average) contribution for all diagrams using
admissible excitor combinations and interaction vertices before scaling by the
time-step $\delta\tau$. The latter can be set according to:
\begin{equation}
  \delta \tau = \frac{\epsilon}{\Xi} = \frac{\epsilon}{\gamma} \max(\chi_c)
\end{equation}
where the parameter $\epsilon$ was introduced in Section \ref{sec:walker-less}
and $\chi_c$ is calculated using $B_{ci}^{\textrm{max}}$. Using
$B_{ci}^{\textrm{ave}}$ would set the time-step such that the average
contribution magnitude was $\epsilon$.

The accumulated values for the diagram-specific terms $B_{ci}^{\star}$ guarantee
that this procedure uses information from all valid diagrams ever generated,
whether spawning attempts were successful or not.
As such, it can converge to a stable importance sampling of the wavefunction
with minimal user input.
To avoid certain classes of diagrams being entirely neglected as a result of a single
negligible diagram, the values $\gamma_{ci}$ must be fixed until sufficient information
has been collected. This is achieved by requiring 100 random diagrams of
each type be selected during the calculation before starting the importance
sampling procedure.

\subsection{Truncated excitation generation and computational scaling}
\label{subsec:trunc_excit_scaling}

The application of truncated excitation generation, as introduced in Section
\ref{subsec:importance_sampling_motiv}, follows naturally in this algorithm, and
provides considerable computational benefits.

Most notably, while the time step is still expected to decrease as $\order{N^{-4}}$ due
to the sampling of the bare Hamiltonian, $H_{\mathrm{N}}$, the cost of sampling
larger clusters will rapidly fall. This is due to the number of possible
diagrams for higher excitation level clusters having a scaling lower than
$\order{N^4}$, and so $\alpha_c$ correspondingly falling with system size to
compensate.

As an example, selected clusters of excitation level $l + 2$ only have $\order{1}$ possible
diagrams due to the restriction to connected diagrams contributing to excitation level $l$
or below. The first example discussed in Section~\ref{subsec:importance_sampling_motiv}
corresponds to this case in \ac{CCSD}. To compensate for this, $\alpha_c$ will fall as
$\order{N^{-4}}$ for this combination, as fewer samples are required.
This gives the overall scaling of sampling these higher terms as $\order{N^{2l+4}}$ for this
case.
For clusters of lower excitation level, sampling the cluster expansion will have
reduced computational expense, which will however be offset by a corresponding
increase in the number of connected diagrams these clusters can contribute to.
This differs from prior unlinked approaches, where all Hamiltonian vertices are
sampled for every cluster, so every cluster effectively contributes to
$\order{N^4}$ (possibly disconnected) diagrams. This results in an additional
factor of $\order{N^4}$ in the computational scaling of unlinked approaches
compared to \ac{diagCCMC}.

As such, the overall cost of a \ac{diagCCMC} calculation will scale as
$\order{N^{2l+4}}$ for a fixed errorbar per electron in the absence of any
simplifying properties of the cluster amplitudes. This gives an asymptotic
scaling of $\order{N^8}$ for \ac{CCSD} with a fixed errorbar per electron. The
asymptotic scaling matches that of deterministic unfactorised \ac{CC} theory,
unlike prior stochastic \ac{CC} approaches.\cite{Scott2017}
If a fixed errorbar were instead required, this would lead to a scaling of
$\order{N^{9}}$ for \ac{CCSD} and $\order{N^{2l+5}}$ in general.

In previous work,\cite{Scott2019-ge} we demonstrated that for systems of
noninteracting replicas the memory cost for \ac{diagCCMC} is proportional to the
number of replicas, regardless of truncation level of the theory, and that this
extends to interacting systems provided cluster amplitudes decay sufficiently
rapidly with distance.
Here, we extend this consideration to show that the computational effort will
asymptotically scale as \emph{at most} $\order{N^4}$ in the presence of locality, regardless of
truncation level, for a fixed errorbar per electron.

Our only requirement is that cluster amplitudes be \emph{homogeneous}: the
absolute magnitude of all $\hat{T}_{m}$ is proportional to some measure of system
size, $N$, as is expected to be the case when locality is present, for instance
in insulators over reasonable length scales.
This means that we can sample the contribution of a cluster of $n$ excitors to a
fixed granularity using only $\order{N^n}$ random samples. The linked diagram
theorem then restricts us to clusters containing at most 4 excitors, and ensures
that the number of ``free'', external coupling indices on the Hamiltonian coupling
vertex which must be sampled is at most $4-n$. Sampling each external index
will require $\order{N}$ additional samples of that diagram, so the maximal
scaling to sample a cluster of size $n(\leq 4)$ is
$\order{N^n}\order{N^{4-n}}=\order{N^4}$. This scaling will be reflected in the
number of attempts per unit of imaginary time, $n_{\textrm{a}}\delta\tau^{-1}$ and is
\emph{independent} of the chosen truncation level in the \ac{CC} hierarchy.
In the case of noninteracting replicas, the computational effort per replica,
$n_{\textrm{a}}\delta\tau^{-1}n_{\textrm{replicas}}^{{-1}}$ is expected to scale as
$\order{N^{3}}$, again independently of truncation level.

It is important to note here a benefit of the stochastic approach: that sparsity
and structure within the cluster amplitudes are exploited automatically, but
only if they are present. In the absence of such structure, the result will
still be equivalent to a conventional \ac{CC} calculation.
This is different from local deterministic approaches, which by necessity
neglect nonlocal contributions according to some categorisation. If locality is
not present to the appropriate degree, such approaches will obtain a different
answer from a conventional \ac{CC} calculation on the same system.
While this may seem technical, being able to exploit locality while still
estimating the exact \ac{CC} energy is a considerable benefit.

\section{Data structures and algorithms}

Our diagCCMC algorithm is implemented in a standalone package. The package is
written in the Python programming language, which allows fast prototyping and
experimentation.

A \emph{sparse stochastic array} is the basic data structure. This is used to
store the compressed representation of cluster operators $T_{m}$ of any rank.
We use a Python \emph{dictionary}: an associative key-value
array implemented as a hash table.\cite{Cormen2009-ik} The excitation indices are the keys:
\begin{equation}
  \tamp{a_1a_2\ldots a_k}{i_1i_2\ldots i_k}
  \longmapsto
  \mathtt{\{ ((i_1,i_2,\ldots, i_k), (a_1,a_2,\ldots, a_k)): t \}},
\end{equation}
with cluster amplitudes stored as floating-point numbers. The key is arranged as
a 2-tuple of $k$-tuples: each $k$-tuple representing the hole and particle
indices, respectively. 
This sparse stochastic array is designed to:
\begin{enumerate*}[label=\emph{\alph*})]
\item be exchange symmetry-aware,
\item perform stochastic rounding to a preset threshold, $\Delta$, and
\item enable importance sampling of its elements.
\end{enumerate*}
The keys in the dictionary are sorted in \emph{ascending} order, both in the
hole and particle tuples: this ensures no storage redundancy.
Upon insertion in the data structure, the supplied index is first sorted in
ascending order, while the supplied value is multiplied by the corresponding
parity phase factor, $(-1)^{\sigma}$. The new value is inserted after stochastic rounding:
\begin{equation}\label{eq:stochastic-round}
  |t| < \Delta \Rightarrow 
  t = 
  \begin{cases}
    (-1)^{\sigma}\sgn(t)\Delta\quad&\text{if}\,\, \text{UniformRandom}_{[0, 1]} < \frac{|t|}{\Delta}  \\
    0\quad&\text{if}\,\, \text{UniformRandom}_{[0,1]}>  \frac{|t|}{\Delta}
  \end{cases},
\end{equation}
finally, the $\ell_{1}$ norm of the $T_{m}$ cluster operator is updated
accumulating the new value:
$\mathtt{||T_{m}||_{1} \leftarrow abs(value)}$.
Similarly, upon lookup, the supplied index is first sorted and then looked up into
the dictionary. If present, the returned value accounts for the parity phase factor.
Iteration and various vector-like operations can be implemented on top of
this storage object.
Compressed sparse matrix formats could replace the hash table. However, the
algorithm is not GEMM-driven\cite{Crawford2019-yz} and compressed sparse
representation would be suboptimal for importance sampling. Insertion and
retrieval are the essential operations in our algorithm and they can be
performed on a hash table with $\order{1}$ complexity in the average case.
Furthermore, data needed for importance sampling can be accumulated upon
insertion into the hash table, eliminating the need for complete traversals of
the data structure.

Importance sampling the data in the sparse stochastic array requires
building the corresponding sampling distribution, either using a cumulative
magnitude array or the alias method\cite{Walker1977-ot,Walker1974-ix} with Vose sampler.\cite{Vose1991-io}
If $n$ is the number of elements in the discrete set to sample, the alias method
constructs the distribution in $\order{n}$, while sampling is achieved in $\order{1}$.

The cluster operator $T$ is a collection of sparse stochastic arrays,
indexed on the rank of its constituent excitations. Various vector-like
algebraic operations can be implemented for this data structure, \eg the
calculation of the $\ell_{1}$ and $\ell_{2}$ norms of $T$, in the form of
reductions over the sparse stochastic arrays of the component operators.

Finally, we handle the importance sampling described in Section
\ref{sec:importance-sampling} in a separate data structure: the sampling store.
This data structure computes the
probabilities for the selection of excitor combinations,
$p_{\mathrm{combo}}$,\cite{Scott2017}  and for the selection of a combination--Hamiltonian vertex
pairing in diagram generation, $p_{\textrm{hver}}$, the latter being conditional on the
former.
The sampling store is also responsible for accumulating data needed to update
the sampling distributions, which is further used to determine the time-step
for the next iteration.

We show a high-level overview of diagCCMC in Algorithm~\ref{alg:mc-loop}.
A diagCCMC calculation requires as input:
\begin{itemize}
  \item Molecular integrals in an orthogonal \ac{MO} basis. These are expected in FCIDUMP
    format.\cite{Knowles1989-uq} 
  \item A truncation level for the cluster operator.
  \item The number of steps, $N_{\mathrm{QMC}}$, to perform.
  \item The stochastic granularity, $\Delta$. This defaults to $10^{-4}$ in our
    implementation.
  \item The time-step, $\delta \tau$, which defaults to $0.01$ in our
    implementation.
  \item The preconditioner, which defaults to the identity in our implementation.
\end{itemize}
The \ac{CC} wavefunction is initialised using the MP1 amplitudes, easily
computed from the provided \ac{MO} basis integrals.
Both the representation of the \ac{CC} wavefunction at the current time-step and
its update (the residual) are represented as stochastic sparse arrays, but
only the former will be used for sampling purposes.
We initialise importance sampling with a short trial run to sample all possible
pairings of excitor combinations and Hamiltonian vertices.

Each \ac{MC} cycle starts by updating the sampling distribution for the cluster
operator: we leverage information about selection probabilities for each
Hamiltonian vertex with each excitor combination, accumulated from diagram
generation attempts in the previous cycle, to define an importance sampling
weight for each combination of excitors.
\begin{algorithm}[H]
  \caption{High-level overview of the diagCCMC algorithm. The main Monte Carlo
    loop in the calculation is initialized given a reference single determinant
    and its Hamiltonian matrix elements in FCIDUMP format.\cite{Knowles1989-uq}
    The MP1 amplitudes are used as the initial $\iterate{\vect{t}}{0}$ guess.}\label{alg:mc-loop}
  \begin{algorithmic}[1]
    \Procedure{diagCCMC}{$H$, $\iterate{\vect{t}}{0}$, $N_{\mathrm{QMC}}$}
    \State Initialize sparse stochastic storage for $T$
    \State Bootstrap importance sampling of $T$
    \For{$n < N_{\mathrm{QMC}}$}
    \State Update sampling distribution for $T^{[n]}$ \Comment{See Section \ref{sec:walker-less-even}}
    \State Stochastic propagation \Comment{See Algorithm \ref{alg:prop}}
    \State Compute the energy deterministically: 
    $$\Delta E_{\mathrm{CC}} = \sum_{ai}t_{a}^{i}f_{i}^{a} +
    \frac{1}{4}\sum_{ijab}t_{ab}^{ij} \twoBodya{i}{j}{a}{b} + \frac{1}{2}\sum_{ijab}t_{a}^{i}t_{b}^{j} \twoBodya{i}{j}{a}{b}$$
    \State Update $T^{[n+1]} \leftarrow \omega^{[n]}$ (``annihilation'') \Comment{See Section \ref{sec:precond}}
    \EndFor
    \EndProcedure
  \end{algorithmic}
\end{algorithm}
The stochastic propagation step performs $n_{\mathrm{a}}$ attempts at sampling
the \ac{CC} residual, $\omega^{[n]}$, constructing diagrams on-the-fly and is schematically described in Algorithm~\ref{alg:prop}.
Our current implementation features a process-based parallelization of this step.
Given $p$ helper processes, each available helper performs
$\lceil\frac{n_{\mathrm{a}}}{p} \rceil$ attempts and stores their results in a queue.\cite{multiprocessing}
These are aggregated by the main process, which also takes care of cleaning up
the queue before entering the next \ac{QMC} step in the simulation.
For each attempt, we first obtain a random cluster and accumulate its relevant
contributions to estimators, \eg the energy.
Given the cluster, we sample its diagonal and off-diagonal actions which we term
``death'' and ``spawn'' attempts, respectively, in analogy with existing Fock-space
\ac{QMC} terminology.
The ``death'' step consists of exact evaluation of all components of the
Hamiltonian which result in contributions to the same excitor as was originally
sampled, provided these have not been incorporated into the preconditioner, as
discussed in~\ref{sec:precond}.
This consists of contributions from vertices 1, 2, 4, 5, and 6 in Table~\ref{tab:H-vertices}.
We may sample \ac{EPV} diagrams\cite{Bartlett2007-mz} within ``death''. These
would cancel out exactly in a deterministic evaluation and are thus not stored
into the sparse representation of the cluster operator.

\begin{algorithm}[H]
  \caption{Stochastic propagation}\label{alg:prop}
  \begin{algorithmic}[1]
    \Procedure{Diagrams}{$H$, $T^{[n]}$, $n_{\mathrm{a}}$}
    \For{$i < n_{\mathrm{a}}$}
    \State Obtain random cluster $\tampc{i}\cdots\tampc{l}$ from $T^{[n]}$
    \State Accumulate contribution of selected cluster to estimators
    \State Sample diagonal action of cluster (``death'') 
    \If{``death'' diagram is \textbf{not} \ac{EPV}}
       \State Apply preconditioning \Comment{See Section \ref{sec:precond}}
       \State Store in residual object after stochastic rounding
    \EndIf{}
    \State Sample off-diagonal action of cluster (``spawn'') \Comment{See Algorithm \ref{alg:diagram-generation}}
    \If{``spawn'' successful}
       \State Evaluate diagram 
       \State Apply preconditioning \Comment{See Section \ref{sec:precond}}
       \State Store in residual object after stochastic rounding
    \EndIf
    \State Accumulate ``spawn'' statistics for importance sampling
    \EndFor
    \EndProcedure
  \end{algorithmic}
\end{algorithm}
During the ``spawn'' attempts, on-the-fly diagram generation will occur, as
described in Algorithm \ref{alg:diagram-generation}. Note that this algorithm is
short-circuiting: an unsuccessful random selection at any step will return an empty
diagram and result in the accumulation of a failed attempt.
\begin{algorithm}[H]
  \caption{Diagram generation}\label{alg:diagram-generation}
  \begin{algorithmic}[1]
    \Procedure{Diagram generation}{$\tampc{i}\cdots\tampc{l}$, sampling distribution, $\refd$}
    \State Select Hamiltonian vertex
    \State Select contraction pattern
    \State Select internal indices
    \State Select external indices \Comment{Spin-conservation constraints are enforced}
    \EndProcedure
  \end{algorithmic}
\end{algorithm}
Once a diagram has been selected, its evaluation is done
\emph{deterministically} by applying the algebraic interpretation rules with
modifications described in Section \ref{sec:diagrammatic-sampling}.
The energy is also evaluated deterministically, but note that a stochastic
estimator can also be built during ``death'' and ``spawning'' steps.
Finally, the cluster operator is updated before moving on to the next \ac{MC} cycle, taking into account the preconditioning of the residual, see Section \ref{sec:precond}.

\section{Numerical examples}

To demonstrate the retention of the favourable properties of our approach when
applied to higher excitation levels in systems of multireference character, we consider calculations including up to
quadruple excitations upon \ce{H4}, in a square of side length \SI{1.5}{\angstrom}.
At this geometry, two \ac{RHF} solutions are
degenerate.\cite{Burton2016holomorphic} Any single configuration provides
only a poor representation of the system, while \ac{CCSDTQ} is equivalent to
\ac{FCI}.
Each truncation level has a clearly identifiable difference in energy,
which can be resolved despite stochastic error. We also consider noninteracting
replicas of this system, such that the wavefunction will become a product.

This system, while small, is by no means trivial for a projection-based
approach. Its multireference nature and small gap between the
ground and excited states necessitates projection through over 50 units of
imaginary time to converge to the ground state. The imaginary-time propagation was not preconditioned. While preconditioning can afford taking larger time steps,\cite{Blunt2019-tf} we observed it can lead to an unstable propagation in this particularly challenging case.

The resultant energies are shown in Table~\ref{tab:h4_replica_energies},
demonstrating the size-extensivity of the energies, within stochastic errorbars, for multiple noninteracting replicas.

\begin{table}[htb]
\caption{
Correlation energies (\si{\hartree}) of \ce{H4} and noninteracting replicas systems at different \ac{CC}
truncation levels in a 6-31G basis. Molecular integrals were generated in
FCIDUMP format\cite{Knowles1989-uq} with the \psicode program
package.\cite{Smith2020-nx}
Deterministic energies calculated using MRCC\cite{Kallay2020-og} for a single replica
are \SI[round-mode=places, round-precision=5]{-0.167460564}{\hartree}, \SI[round-mode=places, round-precision=5]{-0.169980688}{\hartree},
and \SI[round-mode=places, round-precision=5]{-0.162228553}{\hartree}, for \ac{CCSD}, \ac{CCSDT}, and \ac{CCSDTQ}, respectively.
}\label{tab:h4_replica_energies}
  \centering
  \begin{ruledtabular}
   \begin{tabular}{ c  c  c  c }
    $n_\textrm{replicas}$ & \acs{CCSD} & \acs{CCSDT} & \acs{CCSDTQ} \\
    \hline
     1                  & -0.1678(2)  & -0.1701(2)  & -0.1624(2) \\
     2                  & -0.3353(3)  & -0.3398(3)  & -0.3242(8)\\
     3                  & -0.5022(8)  & -0.5129(4)  & \footnote{Values not computed due to computational constraints.} \\
     4                  & -0.6688(7)  & \footnote{Values not computed due to computational constraints.} & \footnote{Values not computed due to computational constraints.}\\
   \end{tabular}
  \end{ruledtabular}
\end{table}

\begin{figure}[hbt]
  \centering
  \includegraphics[width=\textwidth]{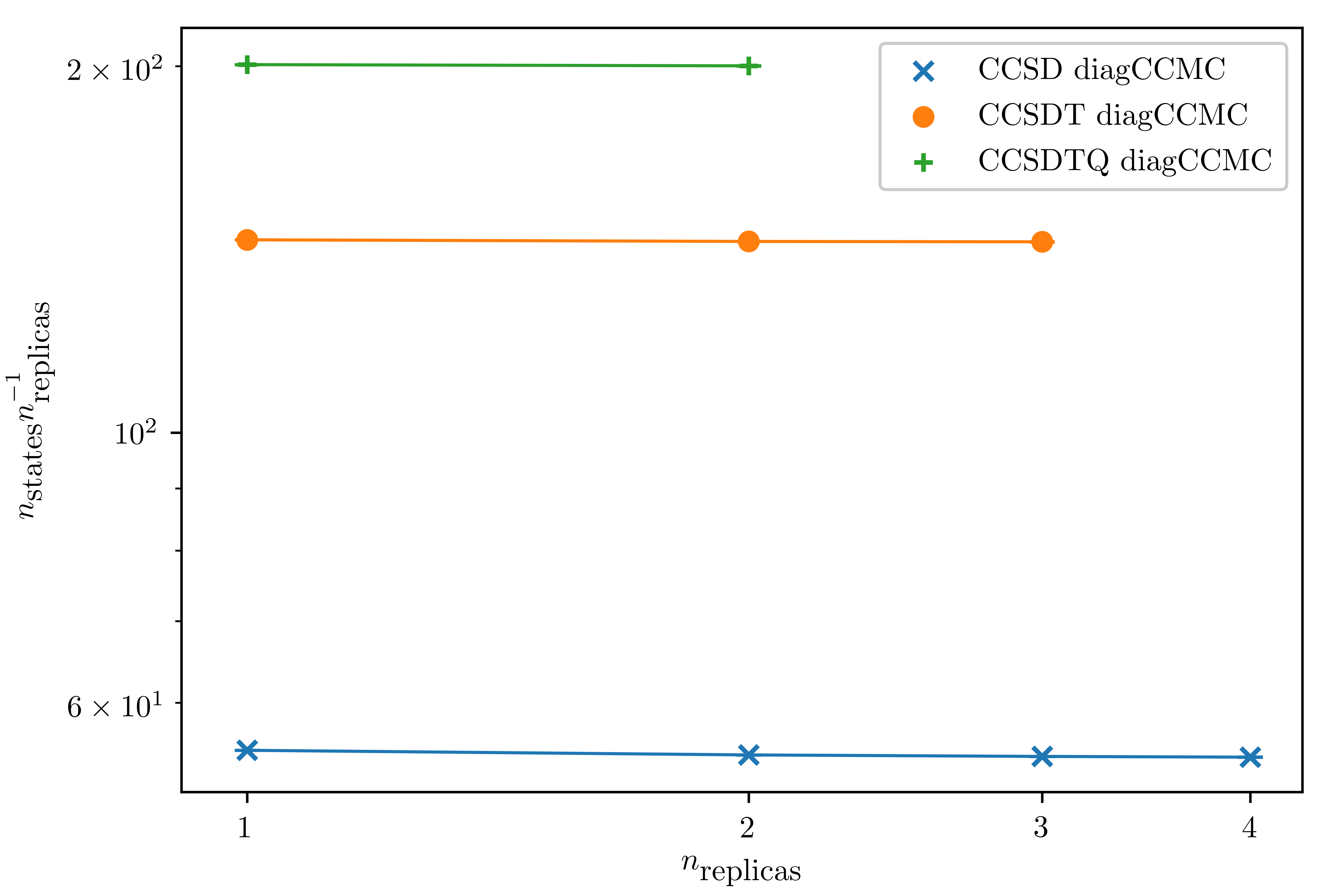}
  \caption{
    $n_{\textrm{states}}$ per replica for noninteracting replicas of \ce{H4}
    in a square geometry of side length \SI{1.5}{\angstrom}, in a 6-31G basis.
    The $n_{\mathrm{states}}$ metric is a measure of the memory cost of the calculation.
    Molecular integrals were generated in FCIDUMP format\cite{Knowles1989-uq} with
    the \psicode program package.\cite{Smith2020-nx}
    Missing points were not computed due to computational constraints.
}
  \label{fig:h4_replicas_nstates}
\end{figure}

The memory cost per replica, as measured by the $n_{\textrm{states}}$ metric, is shown in Figure~\ref{fig:h4_replicas_nstates}. The $\order{N}$ asymptotic scaling for noninteracting systems was already observed in reference \citenum{Scott2019-ge} for noninteracting \ce{Be} replicas systems and is confirmed here also for the \ce{H4} systems. As discussed in Section~\ref{subsec:trunc_excit_scaling}, this is an intrinsic property of the \ac{diagCCMC} algorithm and our results confirm that it is preserved even in cases where the description of the electronic structure is challenging. 

\begin{figure}[hbt]
  \centering
  \includegraphics[width=\textwidth]{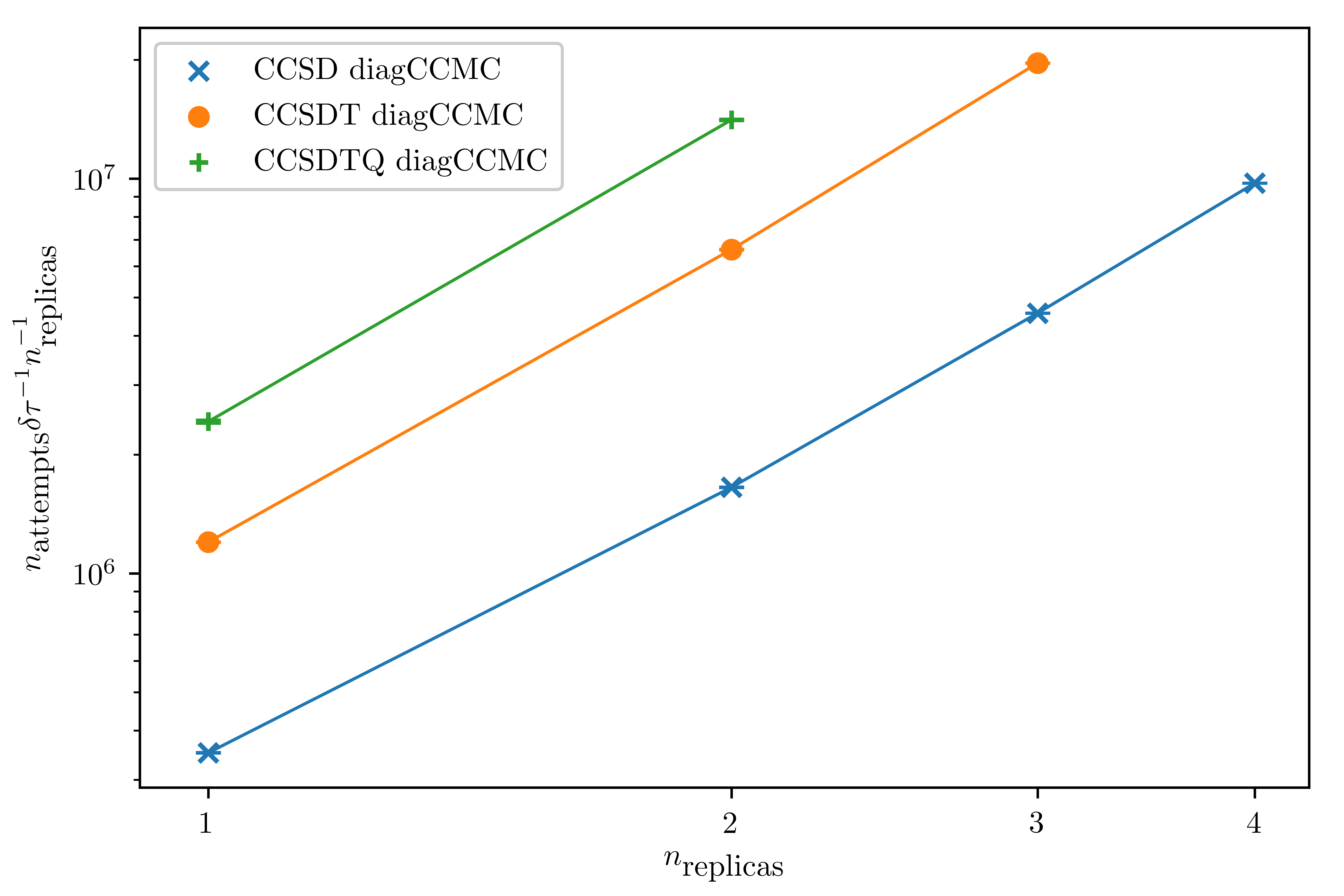}
  \caption{
    Number of stochastic samples performed ($n_{\textrm{a}}$)
    per unit imaginary time per replica for noninteracting
    replicas of \ce{H4} in a square geometry of side length \SI{1.5}{\angstrom},
    in a 6-31G basis.
    This metric is proportional to the CPU cost required to obtain a fixed errorbar per electron.
    Molecular integrals were generated in FCIDUMP format\cite{Knowles1989-uq} with
    the \psicode program package.\cite{Smith2020-nx}
    Missing points were not computed due to computational constraints.
  }
  \label{fig:h4_replicas_nattempts}
\end{figure}

The $n_{\textrm{a}}\delta\tau^{-1}$ metric measures instead the computational requirements per replica and is shown in Figure~\ref{fig:h4_replicas_nattempts}. From the discussion in Section~\ref{subsec:trunc_excit_scaling}, this is expected to scale cubically with 
$n_\textrm{replicas}$.

The results of a log-linear regression analysis 
of the observed $n_{\textrm{a}}\delta\tau^{-1}$ against $n_{\textrm{replicas}}$
are reported in Table~\ref{tab:fits_to_nattempts}. We also include the same analysis on similar data obtained for \ce{Be} in reference \citenum{Scott2019-ge}.
All observed scaling exponents are \emph{below} the expected maximum scaling of $\order{N^3}$. This is not a surprising result: we are not in the asymptotic large-system limit and the highest-scaling contributions will not necessarily dominate the computational cost.

\begin{table}[htb]
\caption{
Scaling exponents and prefactors for the computational scaling of \ac{diagCCMC} calculations with respect to the number of replicas in systems on noninteracting replicas at various \ac{CC} truncation levels. The computational scaling is estimated with the 
$n_{\textrm{a}}\delta\tau^{-1}$ metric discussed in the main text. 
For a dependency $n_{\textrm{a}}\delta\tau^{-1} = c n_{\textrm{replicas}}^{\alpha}$ we fit the linearized model $\ln{(n_{\textrm{a}}\delta\tau^{-1})} = \alpha \ln{n_{\textrm{replicas}}} + \ln{c}$.
Scaling parameters for noninteracting \ce{H4} replicas and noninteracting \ce{Be} replicas are presented. The data for the latter is from reference \citenum{Scott2019-ge} and can be found
at \url{https://doi.org/10.17863/CAM.34952}.
We performed all model fitting using the SciPy package.\cite{Virtanen2020-ud}
}\label{tab:fits_to_nattempts}
\centering
\begin{ruledtabular}
  \begin{tabular}{c  c  c c}
  System & Truncation & $\ln{c}$ & $\alpha$ \\
    \hline
  \multirow{3}{*}{\ce{H4}} & \acs{CCSD}   & $12.73$ & $2.39 \pm0.06$ \\
                           & \acs{CCSDT}  & $13.98$ & $2.53 \pm0.06$ \\
                           & \acs{CCSDTQ} & $14.70$ & $2.54 \pm0.00$ \\
    \hline
  \multirow{3}{*}{\ce{Be}} & \acs{CCSD}   & $12.76$ & $2.75 \pm0.04$ \\
                           & \acs{CCSDT}  & $13.72$ & $2.85 \pm0.03$ \\
                           & \acs{CCSDTQ} & $14.00$ & $2.88 \pm0.02$ \\
  \end{tabular}
\end{ruledtabular}
\end{table}

Finally, we also present correlation energies for the symmetric double dissociation of water in a 6-31G basis, see Table~\ref{tab:h2o_dissoc_energies}. In this system different correlation regimes are in effect along the potential energy surface and it is thus one of the standard benchmarks for correlated methods.\cite{Olsen1996-ab}
\ac{diagCCMC} manages to reproduce values obtained with deterministic approaches at a range of truncation levels along the binding curve.
As is the case for the \ce{H4} calculations presented earlier, the multireference nature of this problem at certain stretched geometries did not allow some of these more challenging calculations to complete.
The \ac{diagCCMC} algorithm is a projection method: despite its intrinsic computational benefits, it still struggles when applied to problems with an ill-defined single reference determinant and/or characterized by a small gap.

\begin{table}[htb]
\caption{
Correlation energies (\si{\hartree}) of the double dissociation of \ce{H2O}  at different \ac{CC}
truncation levels in a 6-31G basis. Molecular integrals were generated in
FCIDUMP format\cite{Knowles1989-uq} with the \psicode program
package.\cite{Smith2020-nx}
Geometries were taken from reference \citenum{Olsen1996-ab}, with $R_{\textrm{e}}=\SI{1.843 45}{\bohr}$.
}\label{tab:h2o_dissoc_energies}
  \centering
  \begin{ruledtabular}
   \begin{tabular}{ c  c  c  c }
    $R_{\textrm{O-H}}/R_{\textrm{e}}$ & \acs{CCSD} & \acs{CCSDT} & \acs{CCSDTQ} \\
    \hline
     1.0                  & -0.13658(5)  & -0.13794(8) & -0.1380(2) \\
     1.5                  & -0.1943(1)  & -0.1997(4) & -0.2008(5) \\
     2.0                  & -0.2906(2)  & -0.3032(3) & \footnote{Value not computed due to computational constraints.} \\
     3.0                  & -0.5315(5)\footnote{The calculation initially converges to the ``canonical" \ac{CCSD} solution, before decaying to a different solution with $\Delta E_\text{CCSD}=\SI{-0.5192(8)}{\hartree}$ after 80 a.u. of imaginary time. This property of the imaginary time propagation has been noted before.\cite{Scott2019-ge}}  & -0.5503(6) & \footnote{Values not computed due to computational constraints.} \\
   \end{tabular}
  \end{ruledtabular}
\end{table}

\section{Conclusions}

We have discussed in detail our new approach for a stochastic solution of the linked
coupled cluster equations, and demonstrated the resulting reduction in computational and memory
costs with system size in the presence of locality. The diagrammatic coupled cluster Monte Carlo
algorithm uses the rigorously order-by-order and term-by-term size-extensive
linked formulation of coupled cluster theory and ensures efficient sampling of
it by on-the-fly construction of coupled cluster diagrams.
The algorithm is made possible by two novel insights:
\begin{enumerate*}[label=\emph{\alph*})]
  \item stochastic compression of multidimensional vectors can be achieved
    \emph{without} invoking walkers and populations, and
  \item the \ac{CC} vector function is an integral, expressible as a finite
    sum of diagrams, that can be computed by Monte Carlo sampling.
\end{enumerate*}
Both insights lead to an algorithm that clarifies how randomness and sampling
can be effectively leveraged to solve the high-dimensional nonlinear \ac{CC}
problem with lower memory footprint and more favorable operation count.
The use of the well-known diagrammatic theoretical framework
clarifies few points of the CCMC methodology, such as the relation of
imaginary-time evolution to iterative solvers and the use of preconditioning.\cite{Blunt2019-tf,Neufeld2020-og}
The representation and evaluation granularity parameters characterize the
diagrammatic approach on a spectrum between fully deterministic and fully
stochastic: the same theoretical framework can accommodate different numerical
approaches.
This paves the way for further cross-adaptation of deterministic and Monte Carlo
techniques.
The approach we have presented uses a na\"{i}ve enumeration of diagrams: the
residual is evaluated in its unfactorised, nonlinear form,\cite{Shavitt2009-mr,Crawford2000} rather than the more
computationally advantageous factorised, quasilinear form.\cite{Stanton1991-dc,Kallay2001-yv}
As such, it exhibits a high operation count, theoretically higher than that of
its deterministic counterpart for a given excitation level if all cluster amplitudes are homogeneous. Deterministically,
one would rather implement a quasilinear factorisation with an optimal space-time
trade-off.\cite{Kallay2001-yv} We are currently investigating this approach.
The use of the diagrammatic expansion also paves the way for a rigorous
derivation of a semistochastic \ac{CC} method, where important residual components are
resolved on-the-fly to machine accuracy, with the remainder only resolved to a preset stochastic
representation granularity.

\section*{Data Availability}

All data and the code used for generation and analysis is freely available at \url{https://doi.org/10.5281/zenodo.3997299}.

\begin{acknowledgments}

C.J.C.S. is grateful to Dr.~George Booth for his current role as Postdoctoral
Research Associate under grant agreement No. 759063 of the European Union’s
Horizon 2020 research and innovation programme.
R.D.R. acknowledges partial support by the Research Council of Norway through
its Centres of Excellence scheme, project number 262695 and through its Mobility
Grant scheme, project number 261873.
T.D.C. was supported by grant CHE-1900420 from the
U.S.~National Science Foundation.
A.J.W.T. is grateful to the
Royal Society for a University Research Fellowship under Grant Nos. UF110161 and
UF160398.
R.D.R. thanks Simen Kvaal (University of Oslo) for pointing out reference \citenum{Jarlebring2014-ke}.

We used the \texttt{goldstone} \LaTeX~package to draw the \acl{CC} diagrams. The
package is available on GitHub: \url{https://github.com/avcopan/styfiles}

\end{acknowledgments}

\appendix

\section{Derivative of the exponential of a parameter-dependent operator}
\label{app:expo-derivative}

Consider an operator $\hat{O}$ dependent on a parameter $\lambda$, its
derivative with respect to $\lambda$ can be obtained as:\cite{Olsen1985-nr}
\begin{equation}
 \begin{aligned}
  \left. \pderiv{}{\lambda}\expT{\hat{O}(\lambda)} \right|_{\lambda = \lambda^{\prime}}
  &=
  \lim_{\delta \rightarrow 0} \frac{1}{\delta} \left[ \expT{\hat{O}(\lambda^{\prime}+\delta)} - \expT{\hat{O}(\lambda^{\prime})} \right] \\
  &\simeq 
  \lim_{\delta \rightarrow 0} \frac{1}{\delta} \left[ \expT{\hat{O}(\lambda^{\prime}) + \dot{\hat{O}}(\lambda^{\prime})\delta} - \expT{\hat{O}(\lambda^{\prime})} \right]
  \end{aligned}
\end{equation}
since to first order in $\delta$ one has $\hat{O}(\lambda^{\prime}+\delta) =
\hat{O}(\lambda^{\prime}) + \dot{\hat{O}}(\lambda^{\prime})\delta$.
The differential $\diff \left[ \expT{\hat{O}(\lambda^{\prime})} \right] =
\expT{\hat{O}(\lambda^{\prime}) + \dot{\hat{O}}(\lambda^{\prime})\delta} -
\expT{\hat{O}(\lambda^{\prime})}$ can be recast as a \ac{BCH} series.
Let us drop the $\lambda^{\prime}$ dependence and rewrite the differential as:
\begin{equation}
  \begin{aligned}
  \diff \left[ \expT{\hat{O}} \right]
&=
  \expT{\hat{O}}
  \left\lbrace \expT{-\hat{O}}\expT{\hat{O} + \dot{\hat{O}}\delta} - 1\right]\rbrace \\
&= 
  \expT{\hat{O}}
  \left\lbrace \expT{-\hat{O} z}\expT{[\hat{O} + \dot{\hat{O}}\delta] z}\right\rbrace_{0}^{1} \\
&=
  \expT{\hat{O}}
  \left\lbrace \int_{0}^{1}\diff z \deriv{}{z} \left[ \expT{-\hat{O} z}\expT{(\hat{O} + \dot{\hat{O}}\delta) z}\right] \right\rbrace
  \end{aligned}
\end{equation}
We can calculate the $z$-derivative as:
\begin{equation}
  \begin{aligned}
 \deriv{}{z} \left[ \expT{-\hat{O} z}\expT{(\hat{O} + \dot{\hat{O}}\delta) z}\right]
&=
  \expT{-\hat{O} z}\lbrace \dot{\hat{O}}\delta \rbrace\expT{(\hat{O} + \dot{\hat{O}}\delta) z} \\
&\simeq
 \delta \expT{-\hat{O} z}\lbrace \dot{\hat{O}} \rbrace\expT{\hat{O} z},
 \end{aligned}
\end{equation}
where in the last step we dropped $O(\delta^{2})$ terms. We then expand the last
term in a \ac{BCH} series:
\begin{alignat}{2}
  \expT{-\hat{O} z}\lbrace \dot{\hat{O}} \rbrace\expT{\hat{O} z}
= \sum_{n \geq 0} \frac{1}{n!} \nestcommR{\dot{\hat{O}}}{\hat{O}z}{n},&\quad
  \nestcommR{\dot{\hat{O}}}{\hat{O}z}{} \stackrel{\textrm{def}}{=} \BCHfirst{\dot{\hat{O}}}{\hat{O}z}.
\end{alignat}
We exchange summation and integration orders and perform the $z$-integration to obtain:
\begin{equation}
  \diff \left[ \expT{\hat{O}} \right]
=
  \delta 
  \expT{\hat{O}}
  \left\lbrace \sum_{n \geq 0} \frac{1}{(n+1)!} \nestcommR{\dot{\hat{O}}}{\hat{O}}{n} \right\rbrace,
\end{equation}
such that the $\lambda$-derivative is:
\begin{equation}\label{eq:exp-op-derivative}
  \pderiv{}{\lambda}\expT{\hat{O}}
=
  \expT{\hat{O}}
 \left\lbrace \sum_{n \geq 0} \frac{1}{(n+1)!} \nestcommR{\dot{\hat{O}}}{\hat{O}}{n} \right\rbrace.
\end{equation}

\bibliography{bibliography}

\end{document}